\documentclass[a4paper]{article}

\usepackage{amsfonts,amsmath,amssymb}
\usepackage{graphicx}
\usepackage{hyperref}
\usepackage[font={small}]{caption}
\usepackage{epstopdf} 
\usepackage{color}
\usepackage[normalem]{ulem}

\def\beq{\begin{equation}}
\def\eeq{\end{equation}}
\def\Rmax{R_{\mathrm{max}}}
\def\Rmin{R_{\mathrm{min}}}
\def\D{\mathrm d}
\def\E{\mathrm e}
\def\GL{\text{GL}}
\def\-{\phantom{-}}

\newcommand{\lowa}[  1]{#1 _{\mathrm a}}
\newcommand{\lowh}[  1]{#1 _{\mathrm h}}
\newcommand{\lowhor}[1]{#1 _{\mathrm{hor}}}

\title{Critical behavior of the \\ black hole / black string transition}
\author{Michael Kalisch\footnote{michael.kalisch@uni-jena.de} , Sebastian Moeckel\footnote{sebastian.moeckel@uni-jena.de} , and Martin Ammon\footnote{martin.ammon@uni-jena.de} \\ \ \\
		\textit{\small{Theoretisch-Physikalisches Institut, Friedrich-Schiller-Universit\" at Jena,}} \\ 
		\textit{\small{Max-Wien-Platz 1, D-07743 Jena, Germany}}}
\date{\today}

\begin{document}

\maketitle

\begin{abstract}

We numerically construct static localized black holes in five and six spacetime dimensions which are solutions to Einstein's vacuum field equations with one compact periodic dimension. In particular, we investigate the critical regime in which the poles of the localized black hole are about to merge. A well adapted multi-domain pseudo-spectral scheme provides us with accurate results and enables us to investigate the phase diagram of those localized solutions within the critical regime, which goes far beyond previous results. We find that in this regime the phase diagram possesses a spiral structure adapting to the one recently found for non-uniform black strings. When approaching the common endpoint of both phases, the behavior of physical quantities is described by complex critical exponents giving rise to a discrete scaling symmetry. The numerically obtained values of the critical exponents agree remarkably well with those derived from the double-cone metric. 

\end{abstract}

\newpage
\tableofcontents

\section{Introduction}
\label{sec:Introduction}

The study of black holes in higher dimensions $D>4$ has become a continuously growing topic over the last decades~\cite{Horowitz:2012nnc}. Considering $D$-dimensional asymptotically flat spacetimes with one periodic spatial dimension, e.g.\ topologically an $\mathbb S^1$, there are at least two different types of solutions: black holes and black strings. These two types can be distinguished by their horizon topology. Black strings wrap the compact dimension, and hence have horizon topology $\mathbb S^{D-3}\times \mathbb S^1$, while black holes are localized on the $\mathbb S^1$ and have  horizon topology $\mathbb S^{D-2}$. 

Uniform black strings in $D$ dimensions may be obtained by adding a compact, periodic dimension (an $\mathbb S^1$) to the solution of a ($D-1$)-dimensional Schwarzschild black hole.\footnote{The generalization of a Schwarzschild solution to higher dimensions is also known as Schwarzschild-Tangherlini solution.} Gregory and Laflamme showed in the seminal papers~\cite{Gregory:1993vy,Gregory:1994bj} that if the size of the compact dimension $L$ is large enough compared to the radius of the Schwarzschild black hole, small perturbations will grow exponentially, hence breaking translational invariance along the $\mathbb S^1$. In other words, assuming $L$ to be fixed, a uniform black string is stable for large masses $M>M_\GL$ and unstable for small masses $M<M_\GL$. Here, $M_\GL$ denotes the mass at the Gregory-Laflamme point, where the uniform black string is marginally stable. 

A few years later it was shown that a new branch of static solutions emanates from the Gregory-Laflamme point. Since these solutions naturally break the translation invariance along the compact periodic dimension, they were called non-uniform black strings. First, they were constructed only in $D=5$ by considering small perturbations around the Gregory-Laflamme point~\cite{Gubser:2001ac}, but later this procedure was adapted to higher dimensions~\cite{Wiseman:2002zc,Sorkin:2004qq}. Beyond the perturbative regime solutions could only be obtained numerically, which was done in a series of papers~\cite{Wiseman:2002zc,Kleihaus:2006ee,Sorkin:2006wp,Headrick:2009pv,Figueras:2012xj,Kalisch:2015via,Kalisch:2016fkm,Dias:2017uyv}, covering the dimensions from $D=5$ up to $D=15$.

Besides black strings there exist black hole solutions with a spherical horizon topology, thus not wrapping the $\mathbb S^1$. These solutions were first discussed in~\cite{Myers:1986rx}. If the size of the event horizon of such a localized black hole\footnote{Sometimes these solutions are also referred to as caged black holes.} is small compared to the size of the compact dimension $L$, the spacetime in the vicinity of the black hole is well approximated by a $D$-dimensional Schwarzschild solution. This fact was used to construct such solutions perturbatively in the regime of small masses~\cite{Harmark:2003yz,Gorbonos:2004uc,Gorbonos:2005px}. Again, for solutions beyond perturbation theory we have to use numerics. This was done in $D=5,6$~\cite{Wiseman:2002ti,Sorkin:2003ka,Kudoh:2003ki,Kudoh:2004hs,Headrick:2009pv} and very recently in $D=10$~\cite{Dias:2017uyv}.

Already in 2002, when numerical data for both the non-uniform black strings and the localized black holes was rare, Kol conjectured that both branches merge into a topology-changing singular transit solution~\cite{Kol:2002xz}. Moving along the localized black hole branch this conjecture implies that the black hole will spread over the $\mathbb S^1$ until its poles merge and the compact dimension is completely wrapped. From the non-uniform black string side the transition is approached when the radius of the string at a certain point on the $\mathbb S^1$ shrinks to zero and the horizon pinches off. A central role in Kol's conjecture plays the so called double-cone metric, which is supposed to be a local model of the transit solution at the critical point, where the merger or the pinch-off happens (depending on from which side the transition is approached). 

All numerical results mentioned above are in accordance with Kol's conjecture, though all implementations break down before the transition is reached. Furthermore, references~\cite{Kol:2003ja,Sorkin:2006wp} provided numerical evidence that the spacetime of non-uniform black strings in $D=6$ dimensions may locally approach the double-cone metric in vicinity of the critical point. 

Recently, non-uniform solutions were constructed far beyond previous results~\cite{Kalisch:2016fkm} and it was possible to investigate the regime of non-uniform black strings very close to the transition. Indeed the authors provided further evidence that the horizon of highly non-uniform black strings in $D=5,6$ converges towards an envelope close to the critical point, exactly described by the double-cone metric. Moreover, in this regime the associated thermodynamic quantities begin to oscillate with a rapidly decreasing amplitude. This led to typical spiraling behavior. In fact, evidence was presented in favor of a distorted logarithmic spiral, implying that not only its amplitude shrinks exponentially with each turn, but also its frequency. Consequently, there may be infinitely many oscillations before the pinch-off is reached.  

In the present work we complement the results of~\cite{Kalisch:2016fkm} by constructing localized black holes in $D=5$ and $D=6$ dimensions. We are particularly interested in the regime, where the poles of the horizon are about to merge on the $\mathbb S^1$. We provide evidence that the associated thermodynamic quantities display similar behavior as in the non-uniform black string case. In particular, we find that these quantities are oscillating around their final values close to the critical solution. The corresponding spiral curves adapt extremely well to the ones of the non-uniform black string branch. 

With the data of both branches at hand, for the first time, we are able to extract critical exponents of the thermodynamic quantities close to the localized black hole/non-uniform black string transition.\footnote{Reference~\cite{Sorkin:2006wp} also tried to find critical behavior of non-uniform black strings but with hindsight the data was not close enough to the transition to observe the scaling.} Moreover, the corresponding critical exponents are in remarkable agreement with the values that Kol derived from the double-cone metric~\cite{Kol:2002xz,Kol:2005vy}.  
 
The paper is structured as follows: We introduce the physical setup for localized black holes in section~\ref{sec:Physical_and_numerical_setup}. In addition, we outline the numerical implementation which is appropriate to obtain highly accurate solutions even close to the transition. In this critical regime the numerics are highly demanding, since the solution approaches a curvature singularity. The heart of our numerical implementation is a multi-domain pseudo-spectral method. In section~\ref{sec:Results} we present our main results and compare them to the non-uniform black string data from~\cite{Kalisch:2016fkm}. We conclude in section~\ref{sec:Conclusions}. Moreover, we provide supplementary material in appendix~\ref{appendix:sec:Details_on_the_numerical_implementation} concerning the numerical implementation. Finally, we briefly review the double-cone metric in appendix~\ref{appendix:sec:Short_review_of_the_double_cone_metric}.

\section{Physical and numerical setup}
\label{sec:Physical_and_numerical_setup}

In this paper we study solutions of Einstein's vacuum field equations
\beq
	R_{\mu\nu} = 0
	\label{eq:Einstein_equations}
\eeq
in $D$ dimensions approaching asymptotically $\mathbb R^{D-3,1} \times\mathbb S^1$ at spatial infinity. In other words, one of the dimensions is compactified to a circle of length $L$. One solution with the correct asymptotics is flat spacetime, given by the following line element
\beq
	\D s^2_{\mathrm{flat}} = -\D t^2 + \D x^2 + \D y^2 + x^2 \D \Omega ^2_{D-3} \, .
	\label{eq:asymptotic_metric}
\eeq
The coordinate $y$ represents the compact dimension. We choose $y\in [-L/2,L/2]$ with period $L$. The $D-2$ infinitely extended spatial dimensions are expressed in spherical coordinates with $x\in [0,\infty ]$ denoting the radial direction and $\D \Omega^2_{D-3}$ denoting the line element of a unit $(D-3)$-sphere.

In the following we construct localized black hole solutions approaching the flat metric~\eqref{eq:asymptotic_metric} in the asymptotic limit $x\to \infty$. We simplify the problem as follows: First, we assume spherical symmetry of the spacetime with respect to the spatially extended dimensions, hence giving rise to the $(D-3)$-sphere discussed above. Second, we restrict ourselves to static solutions. Due to these two assumptions, our problem is effectively two-dimensional with non-trivial behavior in $x$- and $y$-direction. Third, we assume mirror symmetry of the spacetime with respect to the $y=0$ axis. Consequently, due to the periodicity in $y$-direction, we also have mirror symmetry with respect to $y=L/2$. 

\begin{figure}
	\centering
	\includegraphics[scale=1]{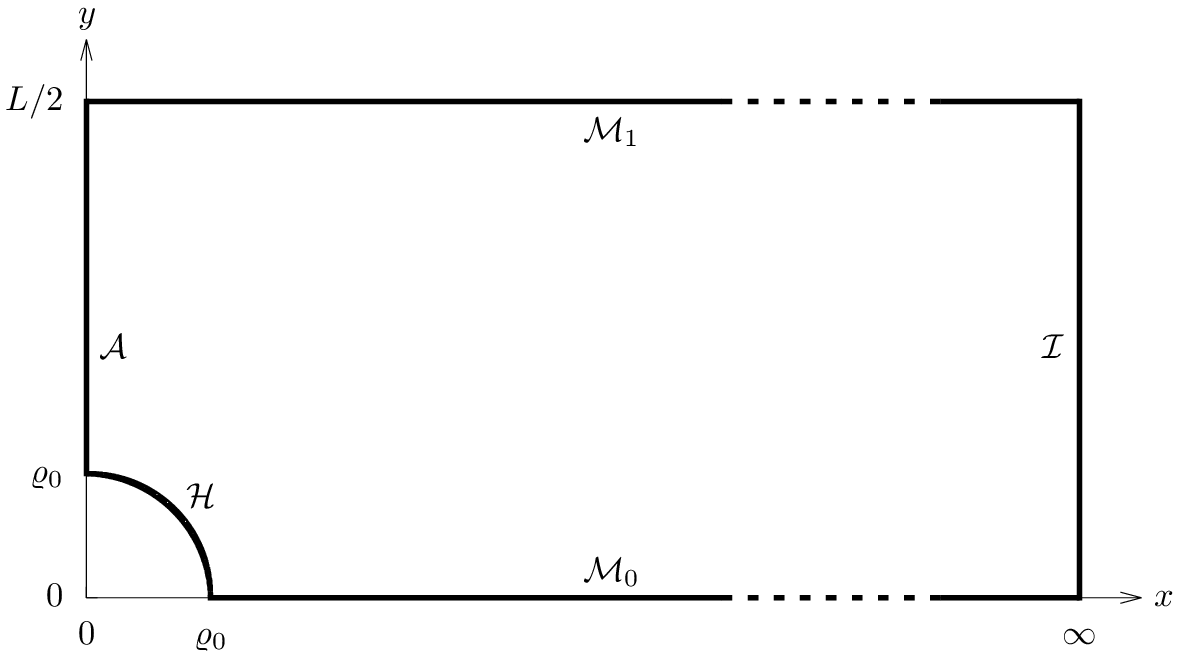}
	\caption{Sketch of the domain of integration with the following boundaries: horizon $\mathcal H$, exposed axis of spherical symmetry $\mathcal A$, lower mirror boundary $\mathcal M_0$, upper mirror boundary $\mathcal M_1$, and asymptotic boundary $\mathcal I$.}
	\label{fig:int_domain_bare}
\end{figure}

Localized black hole solutions have a spherical horizon topology, that is $\mathbb S^{D-2}$. Therefore, the horizon of such objects does not wrap the compact dimension. Let us place the center of the black hole at the origin of the $x$-$y$ plane. It is possible to choose coordinates such that the horizon has a spherical shape with radius $\varrho _0 <L/2$. The corresponding domain of integration is depicted in figure~\ref{fig:int_domain_bare} and contains five boundaries:\footnote{As a consequence of the mirror symmetry with respect to the $y=0$ axis we neglect the lower half of the domain of integration (where $-L/2\leq y\leq 0$) for convenience.}
\begin{itemize}
	\item the horizon $\mathcal H = \{ (x,y)\colon x\geq 0 \, ,~ y\geq 0 \, ,~ x^2+y^2=\varrho _0^2 \}$,
	\item the exposed axis of symmetry $\mathcal A = \{ (x,y)\colon x = 0 \, ,~ \varrho _0 \leq y \leq L/2 \}$,
	\item the lower mirror boundary $\mathcal M_0 = \{ (x,y)\colon x\geq \varrho _0 \, ,~ y = 0 \}$,
	\item the upper mirror boundary $\mathcal M_1 = \{ (x,y)\colon x\geq  0 \, ,~ y = L/2 \}$,
	\item the asymptotic boundary $\mathcal I= \{ (x,y)\colon x\to \infty \, ,~ 0\leq y \leq L/2 \}$.
\end{itemize}
In addition we denote the $y=0$ plane as the equatorial plane and the point $(x,y)=(0,\varrho _0)$ as the north pole of the horizon. Likewise, we refer to $(x,y)=(0,-\varrho _0)$ as the south pole.

We proceed as follows in the remaining part of the section: in section~\ref{subsec:Metric_ansaetze_and_boundary_conditions} we discuss two different coordinate charts, which are adapted to different regions of the domain of integration. The overall numerical scheme is outlined in section~\ref{subsec:Numerical_implementation}, with more details provided in appendix~\ref{appendix:sec:Details_on_the_numerical_implementation}. Finally, we define the relevant physical quantities in section~\ref{subsec:Physical_quantities}.

\subsection{Metric ans\"atze and boundary conditions}
\label{subsec:Metric_ansaetze_and_boundary_conditions}

The integration domain consists of five boundaries as depicted in figure~\ref{fig:int_domain_bare}. It is possible to cover the domain of integration with just a single coordinate chart, for example by constructing appropriate coordinate transformations, as done in~\cite{Kudoh:2003ki}. However, note that such transformations have singular points and due to their complexity the resulting equations of motion will be lengthy.

Instead, we follow the approach of~\cite{Headrick:2009pv} (see also~\cite{Sorkin:2003ka}) and introduce two different coordinate charts: one chart which is adapted to the asymptotic region and another chart which is adapted to the near horizon region.

\ \\ 
\textbf{The asymptotic chart} \nopagebreak \\ 
In order to describe the asymptotic behavior of the metric it is convenient to use the $x$-$y$ coordinates as evident from the line element~\eqref{eq:asymptotic_metric}. The most general ansatz which incorporates the required symmetries reads
\beq
	\D s^2 = - \lowa T \D t^2 + \lowa A \D x^2 + \lowa B \D y^2 + 2 \lowa F \D x \D y +  x^2 \lowa S \D \Omega ^2_{D-3} \, .
	\label{eq:asymptotic_chart}
\eeq
The five metric functions $\lowa T$, $\lowa A$, $\lowa B$, $\lowa F$ and $\lowa S$ depend on $x$ and $y$. We impose the following boundary conditions for these functions on $\mathcal I$, $\mathcal A$, $\mathcal M_0$ and $\mathcal M_1$. 
\begin{itemize}
	\item The asymptotic boundary $\mathcal I$ ($x\to\infty$):\\	
		  In order to approach the flat spacetime~\eqref{eq:asymptotic_metric} asymptotically the metric functions have to satisfy
		  \beq
		  	\lowa T = \lowa A = \lowa B = \lowa S = 1 \quad \mathrm{and} \quad \lowa F = 0 \, .
		  	\label{eq:BCa_asymptotics}
		  \eeq
	\item The exposed axis $\mathcal A$ ($x=0$): \\
		  At this boundary the metric degenerates. Nevertheless, the geometry has to be regular, i.e.\ there is no deficit angle at $x=0$. Therefore we impose
		  \beq
		  	\lowa A = \lowa S \, ,  \quad  \lowa F = 0 \quad \mathrm{and} \quad \frac{\partial \lowa T}{\partial x} = \frac{\partial \lowa A}{\partial x} = \frac{\partial \lowa B}{\partial x} = \frac{\partial \lowa S}{\partial x} = 0  \, .
		  	\label{eq:BCa_axis}
		  \eeq		  
	\item The mirror boundaries $\mathcal M_0$ ($y=0$) and $\mathcal M_1$ ($y=L/2$): \\
		  Since the metric should be symmetric with respect to those two boundaries we demand
		  \beq
		  	\frac{\partial \lowa T}{\partial y} = \frac{\partial \lowa A}{\partial y} = \frac{\partial \lowa B}{\partial y} = \frac{\partial \lowa S}{\partial y} = 0  \quad \mathrm{and} \quad \lowa F = 0 \, .
		  	\label{eq:BCa_mirror}
		  \eeq
\end{itemize}
Note that the flat metric given by the line element~\eqref{eq:asymptotic_metric} already satisfies the conditions above, a fact which will become important later. However, the flat metric obviously does not contain a horizon, where we have to impose  additional conditions. Hence, we introduce another coordinate chart which is suitable for the horizon geometry.

\ \\
\textbf{The near horizon chart} \nopagebreak \\
The horizon is given as a circular contour in the $x$-$y$ plane as depicted in figure~\ref{fig:int_domain_bare}. Therefore, it is convenient to use polar coordinates in the near horizon geometry
\beq
	x =  \varrho \sin \varphi \, , \quad  y = \varrho \cos \varphi \, . 
	\label{eq:polar_coordinates}
\eeq
The metric reads in these coordinates
\beq
	\D s^2 = - \lowh T \D t^2 + \lowh A \D \varrho ^2 + \varrho ^2 \lowh B \D \varphi ^2 + 2 \varrho \lowh F \D \varrho \D \varphi + \varrho ^2 \sin ^2\varphi \, \lowh S \D \Omega ^2_{D-3} \, ,
	\label{eq:horizon_chart}
\eeq
where $\lowh T$, $\lowh A$, $\lowh B$, $\lowh F$ and $\lowh S$ are functions of $\varrho$ and $\varphi$. Comparing equations~\eqref{eq:asymptotic_chart} and~\eqref{eq:horizon_chart} we conclude that $\lowh T = \lowa T$ and $\lowh S = \lowa S$, and that the functions $\lowh A$, $\lowh B$ and $\lowh F$ are linearly connected to the functions $\lowa A$, $\lowa B$ and $\lowa F$. We rewrite the function $\lowh T$ as
\beq
	\lowh T = \kappa ^2 (\varrho -\varrho _0)^2 \lowh{\tilde T}
	\label{eq:funcT}
\eeq
where $\lowh{\tilde T}$ is regular at the horizon. This ensures that the black hole horizon is located at $\varrho =\varrho _0$. The relevant boundary conditions in the horizon chart are listed below.
\begin{itemize}
	\item The horizon boundary $\mathcal H$ ($\varrho =\varrho _0$): \\
		  The metric has to be regular at the horizon and its surface gravity is given by $\kappa$. This leads to the following conditions
		  \beq
		  	\lowh{\tilde T} = \lowh A \, , \quad \frac{\partial \lowh{\tilde T}}{\partial \varrho} = \frac{\partial \lowh A}{\partial \varrho} =  \frac{\partial }{\partial \varrho} (\varrho ^2\lowh B) = \frac{\partial }{\partial \varrho} (\varrho ^2\lowh S) = 0 \quad \text{and} \quad \lowh F = 0 \, .
		  	\label{eq:BCh_horizon}
		  \eeq
	\item The exposed axis $\mathcal A$ ($\varphi =0$): \\
		  Regularity of the metric requires that
		  \beq
		  	\lowh B = \lowh S \, , \quad \quad \frac{\partial \lowh{\tilde T}}{\partial \varphi} = \frac{\partial \lowh A}{\partial \varphi} = \frac{\partial \lowh{B}}{\partial \varphi} = \frac{\partial \lowh S}{\partial \varphi} = 0 \quad \mathrm{and} \quad \lowh F = 0 \, .
		  	\label{eq:BCh_axis}
		  \eeq
	\item The mirror boundaries $\mathcal M_0$ ($\varphi =\pi /2$) and $\mathcal M_1$ ($\varrho \cos \varphi = L/2$): \\
		  It is straightforward to translate the conditions~\eqref{eq:BCa_mirror} into conditions for the functions defined in the near horizon chart.		  
\end{itemize}

\subsection{Numerical implementation}
\label{subsec:Numerical_implementation}

Early numerical works approached the problem by solving Einstein's vacuum field equations~\eqref{eq:Einstein_equations} directly~\cite{Sorkin:2003ka,Kudoh:2003ki,Kudoh:2004hs}. Instead, we employ the well-established DeTurck method~\cite{Headrick:2009pv,Figueras:2011va} (see~\cite{Wiseman:2011by,Dias:2015nua} for reviews). In particular, we solve the Einstein-DeTurck equations (sometimes referred to as  generalized harmonic equations)
\beq
	R_{\mu\nu} - \nabla _{(\mu}\xi_{\nu )} = 0 \, ,
	\label{eq:Einstein-DeTurck_equations}
\eeq
where $R_{\mu\nu}$ and $\nabla_\mu$ are the Ricci tensor and the covariant derivative associated with the target metric $g_{\mu\nu}$. The DeTurck vector field is defined as
\beq
	\xi ^\mu := g^{\alpha\beta} ( \Gamma ^\mu_{\alpha\beta} - \bar\Gamma ^\mu_{\alpha\beta} ) \, .
	\label{eq:DeTurck_vector}
\eeq
Here, $\Gamma^\mu_{\alpha\beta}$ is the Christoffel connection associated with $g_{\mu\nu}$, whereas $\bar\Gamma^\mu_{\alpha\beta}$ is the Christoffel connection associated with a prescribed reference metric $\bar g_{\mu\nu}$. One of the main advantages of the DeTurck method is that the Einstein-DeTurck equations~\eqref{eq:Einstein-DeTurck_equations} are strictly elliptic, in contrast to the original Einstein's field equations~\eqref{eq:Einstein_equations}.

Suppose that the metric $g_{\mu\nu}$ is a solution to the Einstein-DeTurck equations~\eqref{eq:Einstein-DeTurck_equations} and in addition the DeTurck vector field vanishes. Then $g_{\mu\nu}$ solves Einstein's vacuum field equations~\eqref{eq:Einstein_equations}. In particular, we have to ensure that the reference metric satisfies the same boundary conditions as the target metric. This is a necessary condition for a vanishing DeTurck vector field. This condition does not completely rule out the occurrence of so-called Ricci solitons, i.e.\ solutions to~\eqref{eq:Einstein-DeTurck_equations} which do not solve~\eqref{eq:Einstein_equations}. However, as shown in~\cite{Figueras:2011va}, Ricci solitons do not exist in the static case considered here. Nonetheless, in practice we solve~\eqref{eq:Einstein-DeTurck_equations} numerically and we explicitly check afterwards if the DeTurck vector field decreases to reasonably small values.

In order to implement the DeTurck method, we have to find an appropriate reference metric $\bar g_{\mu\nu}$ which is consistent with the boundary conditions. For this purpose we follow the strategy outlined in~\cite{Headrick:2009pv} with some minor modifications. We use the flat metric~\eqref{eq:asymptotic_metric} as the reference metric outside the ball of radius $\varrho_1$ centered in the origin of the $x$-$y$ plane, where $\varrho_1$ is chosen such that $\varrho _0 < \varrho _1 < L/2$. As stated beforehand, this metric satisfies the conditions on the outer boundaries. Inside this ball we have to use a reference metric that contains a horizon at $\varrho = \varrho _0$ with surface gravity $\kappa$. A natural candidate for such a spacetime is the $D$-dimensional Schwarzschild metric. However, this metric itself is not appropriate, since it does not match with the flat metric at $\varrho = \varrho _1$. Instead we modify the Schwarzschild metric appropriately. We refer the reader to appendix~\ref{appendix:subsec:Choice_of_reference_metric_functions} for a detailed description of the modification.  

Note that the surface gravity of the reference metric has to be identical to the surface gravity of the desired target metric, when considering the ansatz~\eqref{eq:horizon_chart} and the redefinition~\eqref{eq:funcT}. Consequently, by varying $\kappa$ we construct physically inequivalent localized black hole solutions. In contrast, the parameters $\varrho _0$ and $\varrho _1$ can only change the gauge but they have no influence on physical properties of the solution, provided that they are chosen reasonably. 

Having a reference metric at hand, we solve the Einstein-deTurck equations in the two different charts, that are discussed in the previous section. For this purpose, the domain of integration has to be divided in an asymptotic region, where the asymptotic chart is appropriate, and a near horizon region for which we use the near horizon chart. For convenience, we choose the boundary between these regions to be the contour $x=L/2$. In order to cover the whole domain of integration on a numerical grid, a coordinate transformation compactifying infinity $\mathcal I$ has to be performed in the asymptotic region. Moreover, we decompose the near horizon region, which has five boundaries, in such a way that each subdomain only has four boundaries. This is beneficial since there exists a coordinate transformation in each subdomain which is non-singular and covers the corresponding subdomain without an overlap to other domains. We further subdivide the integration domain using the special contour $\varrho =\varrho _1$ where the two different reference metrics have to be matched. The resulting basic arrangement of domains is depicted in figure~\ref{fig:int_domain_all_grid}. In appendix~\ref{appendix:subsec:Decompositoin_of_the_domain_of_integration} we give the corresponding coordinate transformations and discuss some further adaptions.
\begin{figure}
	\centering
	\includegraphics[scale=1.0]{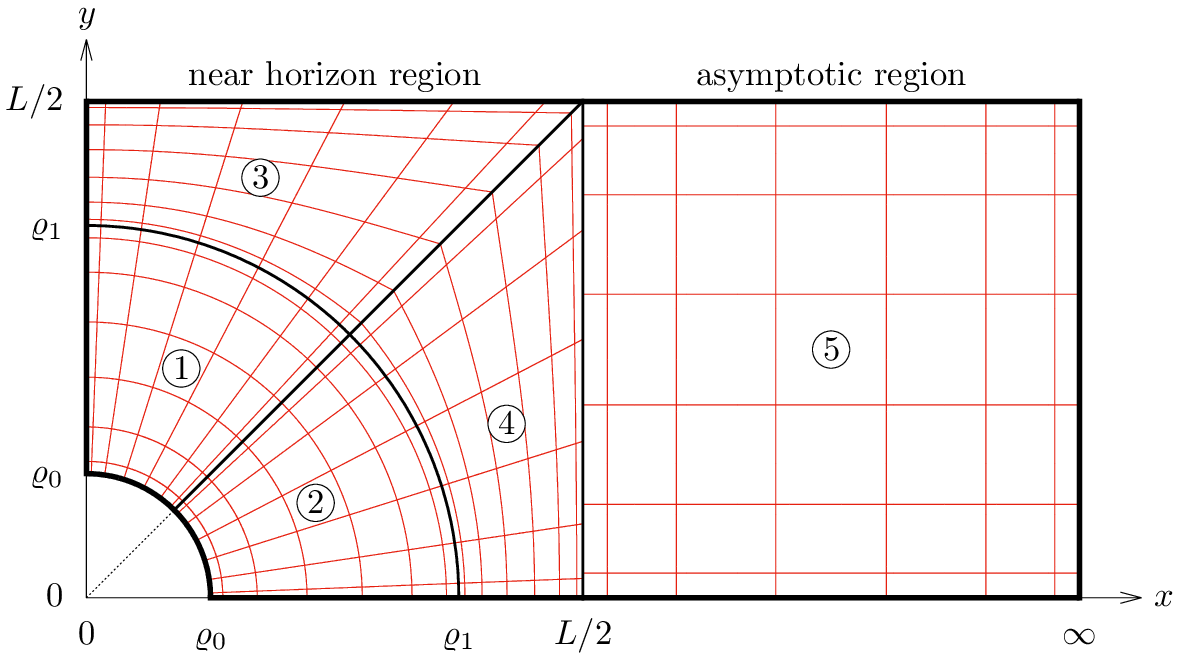}
	\caption{Basic decomposition of the domain of integration and schematic collocation of grid lines. Coordinate charts: we use the asymptotic chart~\eqref{eq:asymptotic_chart} for $x\geq L/2$ and the near horizon chart~\eqref{eq:horizon_chart} for $x\leq L/2$. Reference metric: for $\varrho \geq \varrho _1$ we use flat spacetime~\eqref{eq:asymptotic_metric} as the reference metric, while for $\varrho <\varrho _1$ we use a modified Schwarzschild metric satisfying the appropriate boundary conditions at $\mathcal H$, $\mathcal A$ and $\mathcal M_0$ and matching to the flat reference metric at $\varrho = \varrho _1$, see appendix~\ref{appendix:subsec:Choice_of_reference_metric_functions} for more details.}
	\label{fig:int_domain_all_grid}
\end{figure}

In the numerical implementation we omit one of the boundary conditions at the horizon $\mathcal H$ and at the exposed axis $\mathcal A$, respectively. This is due to the fact that there are six conditions for only five unknowns. In theory we are free to omit any of these conditions~\cite{Dias:2015nua}. At the end, the numerical solution satisfies the missing condition, at least up to  numerical precision. In practice we omit $\partial \lowh{\tilde T}/\partial \varrho=0$ at the horizon $\mathcal H$ and $\partial \lowh S/\partial\varphi=0$ at the exposed axis $\mathcal A$.

Finally, we discretize the resulting Einstein-DeTurck equations and boundary conditions by means of a pseudo-spectral method. The basis of our scheme is the expansion of any function $f\colon [a,b]\to \mathbb{R}$ in terms of Chebyshev polynomials $T_k(y) = \cos [k \arccos (y)]$, where $y\in [-1,1]$ and
\beq
	f(x) \approx \sum _{k=0}^{N-1} c_k T_k\left( \frac{2x-b-a}{b-a} \right) \, .
	\label{eq:spectral_expansion}
\eeq 
In the process, the function values are considered on Lobatto grid points which include the boundaries $a$ and $b$,
\beq
	x_k = (b-a) \sin ^2\left[ \frac{\pi k}{2(N-1)} \right] + a \quad \text{with} \quad k=0,1,\ldots ,N-1  \, .
	\label{eq:Lobatto}
\eeq
We determine the spectral coefficients $c_k$ by utilizing a slightly modified version of the FFTW algorithm~\cite{FFTW05}. We apply the Newton-Raphson method for solving the discretized system describing the collection of non-linear partial differential equations~\eqref{eq:Einstein-DeTurck_equations}. In the several iterative steps of this scheme a linear system involving a Jacobian matrix has to be solved. This is done by means of the so-called BiCGSTAB method~\cite{barrett1994templates} in combination with a preconditioner that utilizes a finite difference representation of the Jacobian. The sparse linear system arising within the preconditional step is solved efficiently (in terms of memory and time consumption) with the help of the SuperLU library~\cite{superlu99,superlu_ug99}.

\subsection{Physical quantities}
\label{subsec:Physical_quantities}

In this section we determine the physical quantities of interest for us. In particular, we are interested in two asymptotic charges and several geometric quantities measured on the horizon $\mathcal H$ or on the axis of symmetry $\mathcal A$. 

\ \\
\textbf{Asymptotic charges} \nopagebreak \\
We already stressed that in the asymptotic limit $x\to\infty$ the metric approaches flat spacetime~\eqref{eq:asymptotic_metric}. In the presence of a black hole the relevant leading order corrections to the metric at infinity are~\cite{Kol:2003if,Harmark:2003dg} 
\beq
	\lowa T \simeq  1 - \frac{c_t}{x^{D-4}} \, , \quad \lowa B \simeq 1 + \frac{c_y}{x^{D-4}} \, .
	\label{eq:asymptotic_corrections}
\eeq
From the two coefficients $c_t$ and $c_y$ we can determine the mass of the black hole 
\beq
	M =  \frac{L \Omega _{D-3}}{16 \pi G_D} \left[ (D-3) c_t - c_y ) \right] \, ,
	\label{eq:mass}
\eeq
as well as the relative tension 
\beq
	n = \frac{c_t - (D-3) c_y}{(D-3) c_t - c_y} \, .
	\label{eq:tension}
\eeq
Here, $\Omega _{D-3}$ represents the surface area of a ($D-3$)-sphere and $G_D$ is the gravitational constant in $D$ dimensions.

\ \\ 
\textbf{Geometric quantities} \nopagebreak \\
According to the laws of black hole thermodynamics, the temperature $T$ of the black hole is proportional to the surface gravity, $T = \kappa /(2\pi )$, and the associated entropy $S$ is related to the horizon area $A$ by $S=A/(4 G_D)$. Note that we do not have to extract the surface gravity $\kappa$ from the numerical solution since we prescribe it by choosing the appropriate near horizon reference metric. We determine the horizon area $A$ within the near horizon chart~\eqref{eq:horizon_chart} by computing
\beq
	A = 2 \varrho _0^{D-2} \Omega _{D-3} \int _0^{\pi /2} \sqrt{\lowh B \lowh S^{D-3}} ( \sin  \varphi )^{D-3} \, \D \varphi \, ,
	\label{eq:horizon_area}
\eeq
where the metric functions are evaluated at $\varrho = \varrho _0$. Furthermore, we characterize the horizon with the following quantities: The horizon areal radius at the equator
\beq
	R_{\text{eq}} = \varrho _0 \sqrt{\lowh S} \quad \text{at} \quad ( \varrho , \varphi ) = ( \varrho _0 , \pi /2 ) \, , 
	\label{eq:Req}
\eeq
and the proper distance from north to south pole along the horizon
\beq
	L_\text{polar} = 2\varrho _0 \int _0^{\pi /2} \sqrt{\lowh B} \, \D \varphi \quad \text{at} \quad \varrho = \varrho _0 \, . 
	\label{eq:Lpolar}
\eeq
The proper distance between the poles along the axis of symmetry
\beq
	L_{\text{axis}} = 2 \int _{\varrho _0}^{L/2} \sqrt{\lowh A} \, \D \varrho \quad \text{at} \quad \varphi = 0 
	\label{eq:Laxis}
\eeq
is useful to distinguish between physically inequivalent solutions.  For $L_{\text{axis}} \to L$ the radius of the black hole vanishes and hence we obtain flat spacetime~\eqref{eq:asymptotic_metric}. Moreover, we approach the transit solution in the limit $L_{\text{axis}} \to 0$.

For the purpose of illustration we embed the ($D-2$)-dimensional surface of the horizon into ($D-1$)-dimensional flat space
\beq
	\D s^2 = \D X^2 + \D Y^2 + X^2 \D \Omega ^2_{D-3} \, .
	\label{eq:embed_metric}
\eeq
Comparing this expression to~\eqref{eq:horizon_chart} we deduce 
\refstepcounter{equation}\label{eq:embedding}
\begin{align}
	X(\varphi ) &= \varrho _0 \sin \varphi \sqrt{\lowh S}  \, , \tag{\theequation a} \label{eq:embedding_a} \\
	Y(\varphi ) &= \int _\varphi ^{\pi /2} \sqrt{\varrho _0^2 \lowh B - (X_{,\tilde\varphi})^2} \, \D \tilde \varphi  \, , \tag{\theequation b} \label{eq:embedding_b} 
\end{align}
where both $X$ and $Y$ are evaluated at the horizon $\varrho =\varrho _0$. We have chosen an arbitrary constant in the integration of~\eqref{eq:embedding_b} such that $Y=0$ at the equator in analogy to the coordinate $y$.

\ \\ 
\textbf{First law} \nopagebreak \\
For static black holes in spacetimes with one compact dimension the first law of black hole thermodynamic reads
\beq
	\D M = T \D S + \frac{n M}{L} \D L \, .
	\label{eq:FirstLaw}
\eeq
Since we keep the asymptotic length of the compact dimension $L$  fixed in our case, the first law reduces to $\D M = T \D S$. There is also an integrated form of the first law~\cite{Kol:2003if,Harmark:2003dg}, known as Smarr's relation, which reads 
\beq
	(D-2) TS = (D-3-n) M  \, .
	\label{eq:Smarr}
\eeq 
Note that equation~\eqref{eq:Smarr} may serve as a non-trivial consistency check for a given numerical solution, since horizon values are related to the asymptotic value $c_t$.\footnote{Note that $c_y$ drops out of the right hand side of Smarr's relation~\eqref{eq:Smarr}.}

\section{Results}
\label{sec:Results}

In this work we use the previously described scheme to construct localized black hole solutions in five and six spacetime dimensions. The rather complicated domain setup allows us to increase the numerical resolutions where it is needed -- in particular near the axis $\mathcal A$, the horizon $\mathcal H$ and the asymptotic boundary $\mathcal I$, see appendix~\ref{appendix:subsec:Decompositoin_of_the_domain_of_integration}. Consequently, we are able to extend the branch of solutions far beyond previous results, while coming extremely close to the transit solution, where a transition to the non-uniform black string branch is supposed to happen. We have to increase the resolution considerably when approaching the transit solution to maintain accuracy. The computing time for a single numerical solution still remains small enough to be carried out on an ordinary PC within a couple of minutes.

First, we show the qualitative behavior of the thermodynamic and geometric quantities in sections~\ref{subsec:Thermodynamics} and~\ref{subsec:Geometry}. In particular, section~\ref{subsec:Geometry} illustrates how the shape of the horizon changes when approaching the transit solution. In section~\ref{subsec:Critical_behavior} we present the main results of this work: We investigate the critical regime in more detail and find that the behavior of different physical quantities is governed by the same complex critical exponents, whose values were already conjectured~\cite{Kol:2002xz,Kol:2005vy}. Due to the complex nature of the components we find evidence for a discrete scaling symmetry. 

Moreover, in this section we compare the localized black hole solutions with the non-uniform black string solutions of~\cite{Kalisch:2016fkm}. In particular, we provide numerical evidence that indeed both branches are merging in the same transit solution. Finally, we discuss the accuracy of the localized black hole solutions obtained in this work in section~\ref{subsec:Accuracy}.

\subsection{Thermodynamics}
\label{subsec:Thermodynamics}

\begin{figure}
	\centering
	\includegraphics[scale=1]{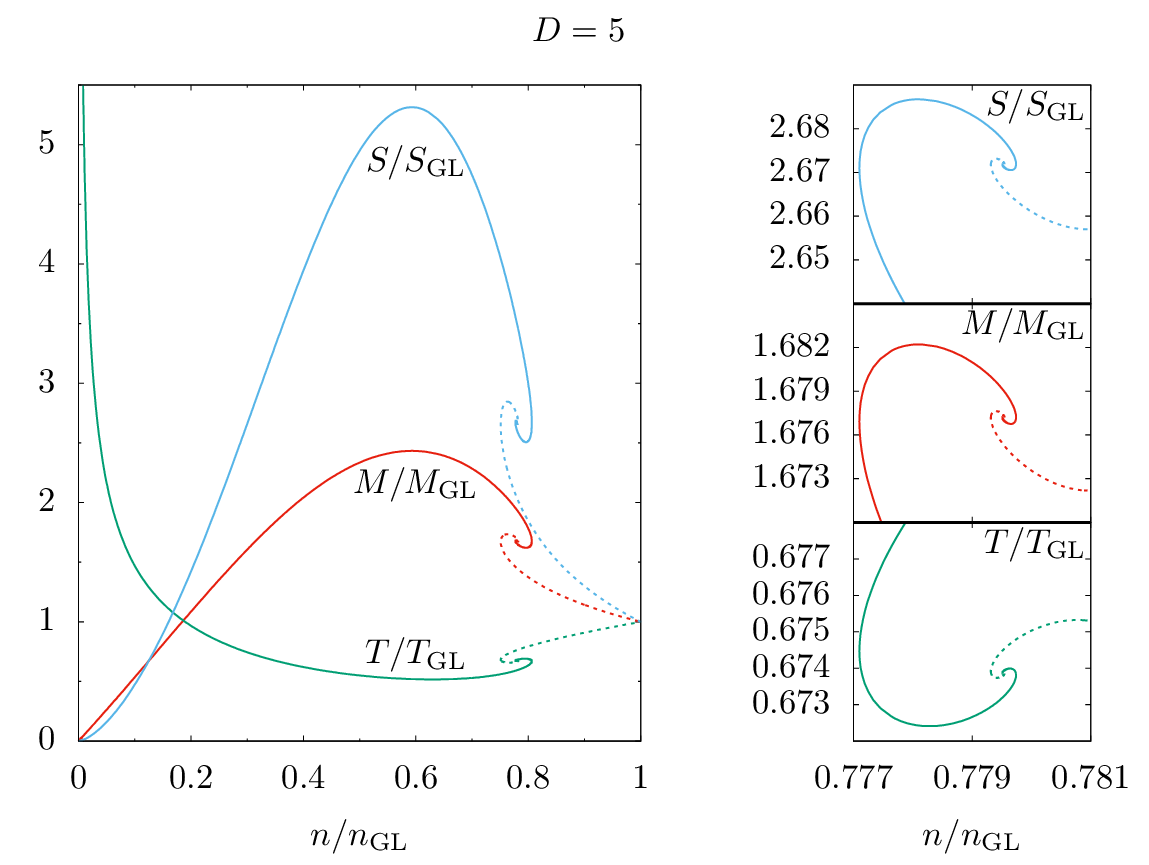} \ 
	\includegraphics[scale=1]{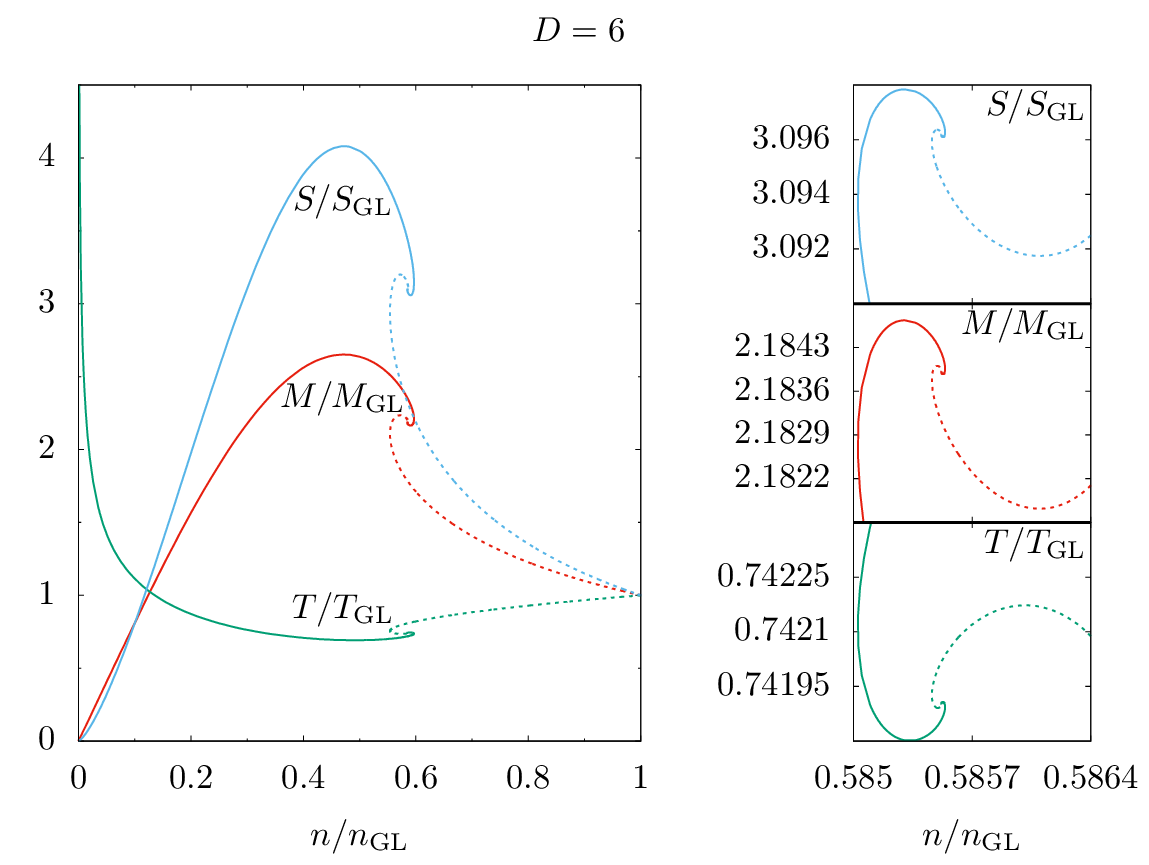}
	\caption{Temperature $T$, entropy $S$ and mass $M$ as a function of the relative tension $n$. The solid lines represent the localized black hole branch while the dashed lines correspond to the non-uniform black strings (uniform black strings have $n/n_\text{GL}=1$). All variables are normalized by their corresponding value of a marginal unstable uniform black string at the Gregory-Laflamme point.  The regions where the curves approach each other are magnified and displayed at the right hand side.}
	\label{fig:Thermo}
\end{figure}

In figure~\ref{fig:Thermo} we use the relative tensions $n$ to parametrize the solutions along both the localized black hole branch and the non-uniform black string branch. We normalize the relative tension and the other thermodynamic quantities by their corresponding values of a marginally stable uniform black string at the Gregory-Laflamme point, indicated by the subscript $\GL$. The parametrization with respect to $n$ is convenient since the $n/n_\GL=1$ line corresponds to the uniform black strings, from which the non-uniform black string branch emanates with values of $n/n_\GL$ smaller than one.\footnote{All uniform black strings have the same tension, i.e.\ $n/n_\GL=1$.} The localized black hole branch starts where $n/n_\GL$ is close to zero and the black hole is very small compared to the size of the compact dimension. By increasing the value of $n/n_\GL$, the size of the localized black hole increases. However, as previous studies~\cite{Kudoh:2004hs,Headrick:2009pv} already pointed out, at some point the horizon area (or entropy) and the mass reach their maxima and then they decrease while heading towards the non-uniform black string branch. 

The present study closes the gap between the localized black holes and black strings, as it is illustrated in figure~\ref{fig:Thermo}. Following the localized black hole branch, we see that the first clearly pronounced turning point of the displayed thermodynamic variables is followed by at least three more. This leads to the spiraling behavior of the curves in figure~\ref{fig:Thermo}. Moreover, these spirals adapt perfectly to those of the non-uniform black string branch. At each twist of the spirals the two phases approach each other. This strongly supports the expectation that both branches eventually end at the same point in the phase diagram. 

In~\cite{Kalisch:2016fkm} it was argued for the non-uniform black string phase that the extend of the spirals shrinks exponentially with each turn (like for a logarithmic spiral), which led to the conjecture that there will be infinitely many turns before the endpoint is reached. This will be confirmed in section~\ref{subsec:Critical_behavior}.

\begin{figure}
	\centering
	\includegraphics[scale=0.98]{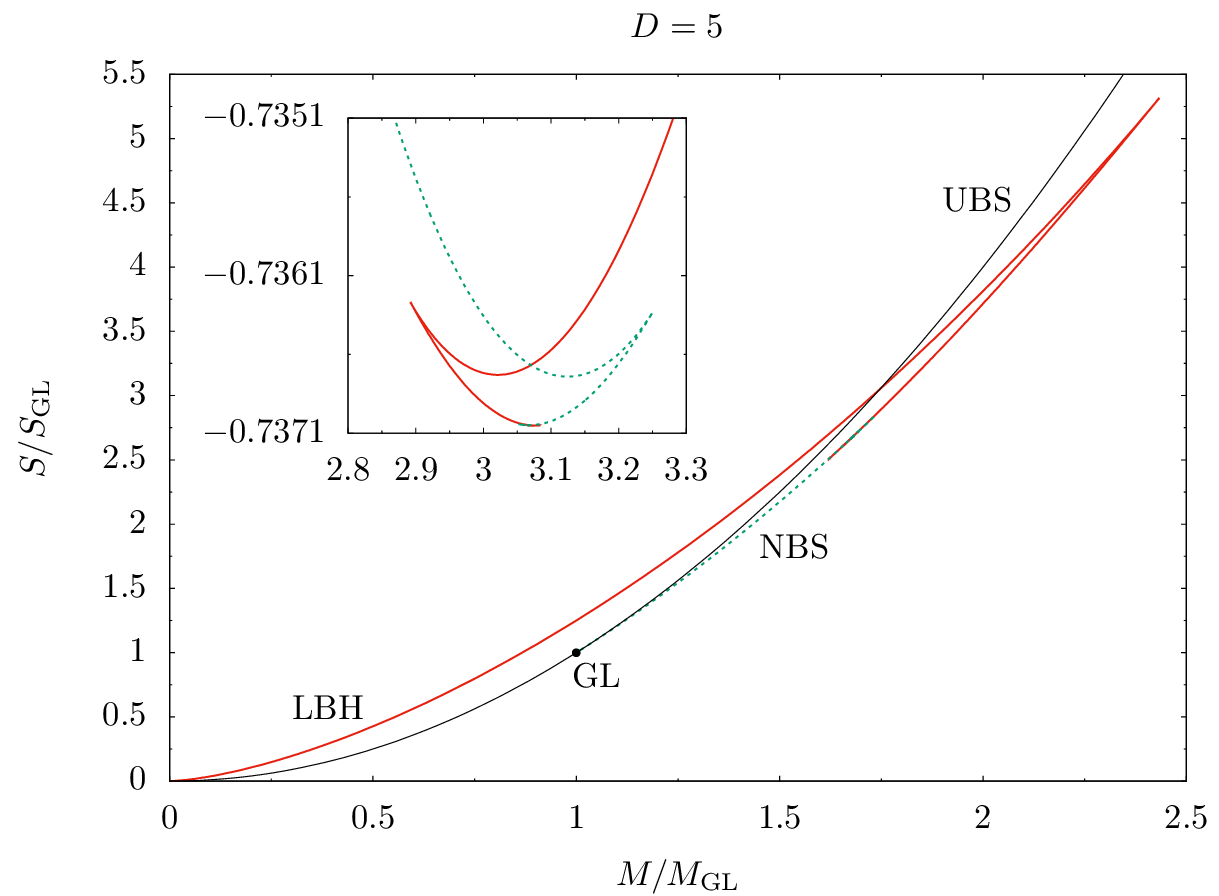} \ 
	\includegraphics[scale=0.98]{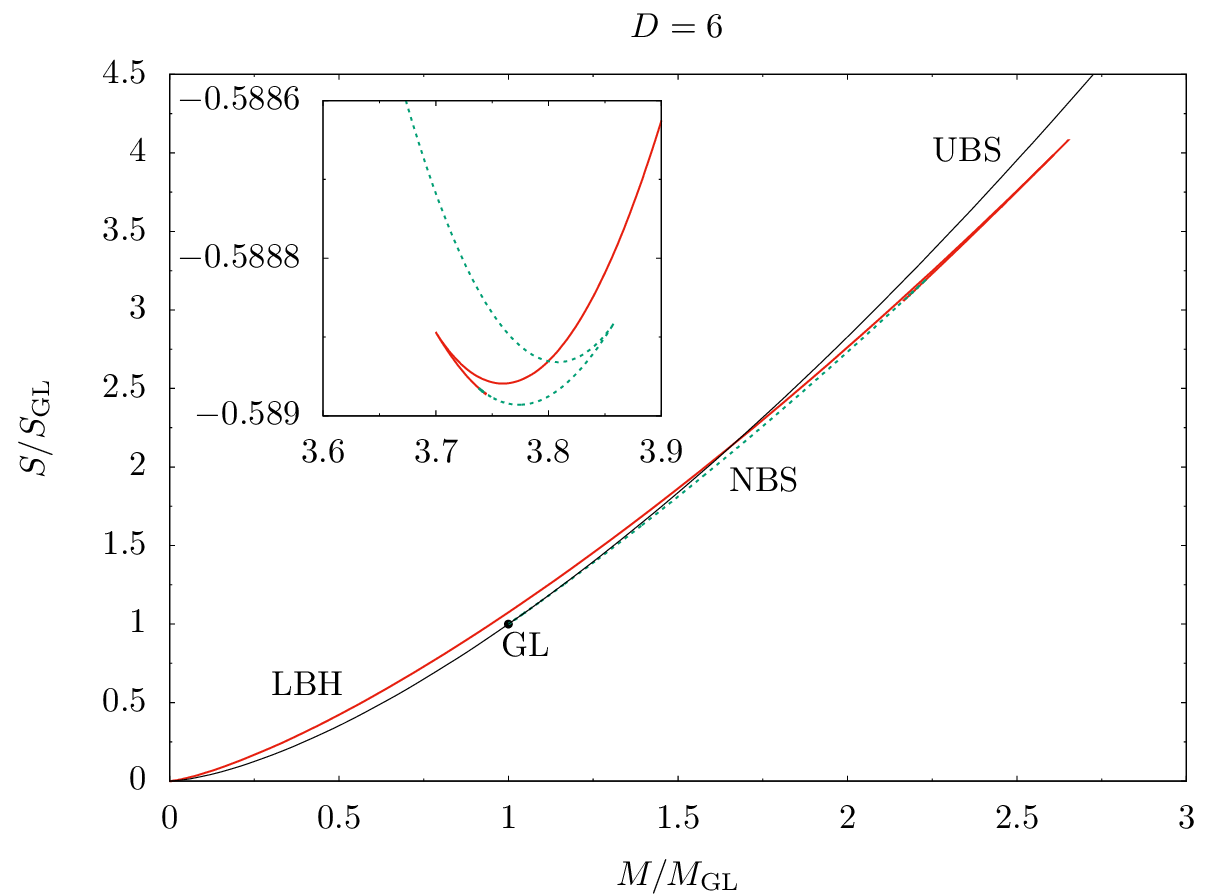}
	\caption{Phase diagram in the microcanonical ensemble: entropy $S$ as a function of the mass $M$. The solid red line represents the localized black hole branch while the dashed green line corresponds to the non-uniform black strings. The thin black line represents the uniform black strings. All variables are normalized by their corresponding value of a marginal unstable uniform black string at the Gregory-Laflamme point. In the inset we display the region around the transit solution (appropriately magnified and rotated which results in the different scales on the axes of the inset).}
	\label{fig:MicroEnsemble}
\end{figure}

For completeness we determine the phase diagram in the microcanonical ensemble. In figure~\ref{fig:MicroEnsemble} we display the entropy $S$ as a function of the mass $M$. The thermodynamically preferred solutions for a given mass $M$ are the ones with largest entropy, i.e.\ for small masses the localized black hole solutions and for large masses the uniform black strings. In the inset of figure~\ref{fig:MicroEnsemble} we magnify the region around the transit solution and show that the localized black holes turn into non-uniform black strings.

\subsection{Geometry}
\label{subsec:Geometry}
\begin{figure}
	\centering
	\includegraphics[scale=1]{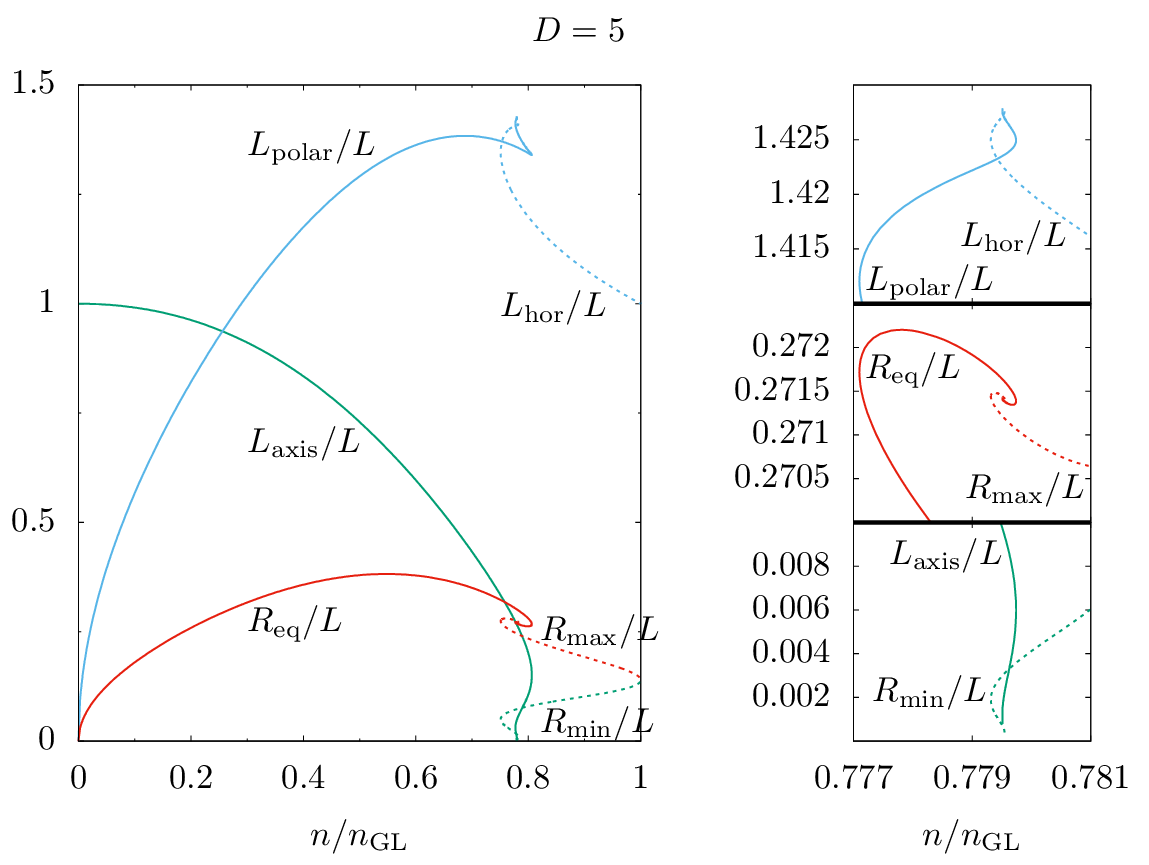} \ 
	\includegraphics[scale=1]{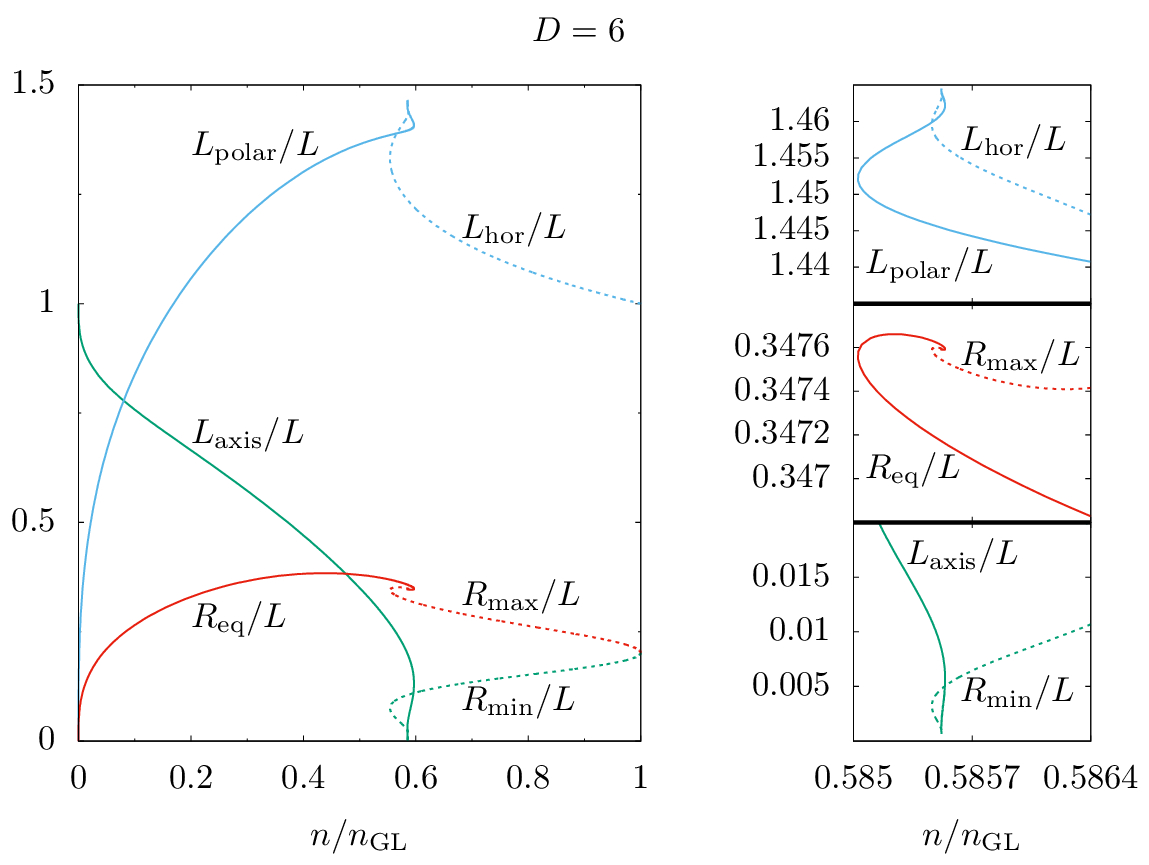}
	\caption{Several geometric quantities as a function of the relative tension $n$. The quantities of the localized black hole branch $L_{\text{polar}}$, $L_{\text{axis}}$ and $R_{\text{eq}}$ are represented by solid lines. For the black strings we display the minimal and maximal horizon areal radius, $R_\text{min}$ and $R_\text{max}$, and the proper length of the horizon along the compact dimension, $L_\text{hor}$, as dashed lines.}
	\label{fig:Geometry}
\end{figure}

In this section we discuss geometrical aspects of the transit solution between non-uniform black strings and localized black holes and we provide additional evidence in favor of the conjectured double-cone metric. Figure~\ref{fig:Geometry} shows the behavior of some geometric quantities as a function of the relative tension $n$. In particular, we display the horizon areal radius at the equator $R_{\text{eq}}$~\eqref{eq:Req}, the inter-polar distance along the axis of spherical symmetry $L_{\text{axis}}$~\eqref{eq:Laxis} and the polar distance along the horizon $L_{\text{polar}}$~\eqref{eq:Lpolar} for the localized black hole solutions. In addition, we also show the corresponding quantities of black strings, i.e.\ the maximal and minimal horizon areal radius on the $\mathbb S^1$, $\Rmax$ and $\Rmin$, and the proper length of the horizon along the $\mathbb S^1$, $L_\text{hor}$.

\begin{figure}
	\centering
	\includegraphics[scale=0.99]{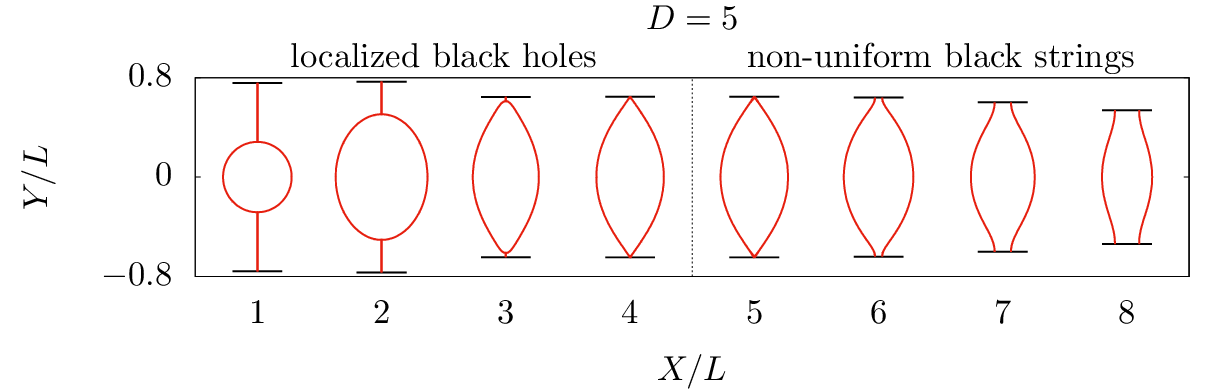} \ \\ \ \\
	\includegraphics[scale=0.99]{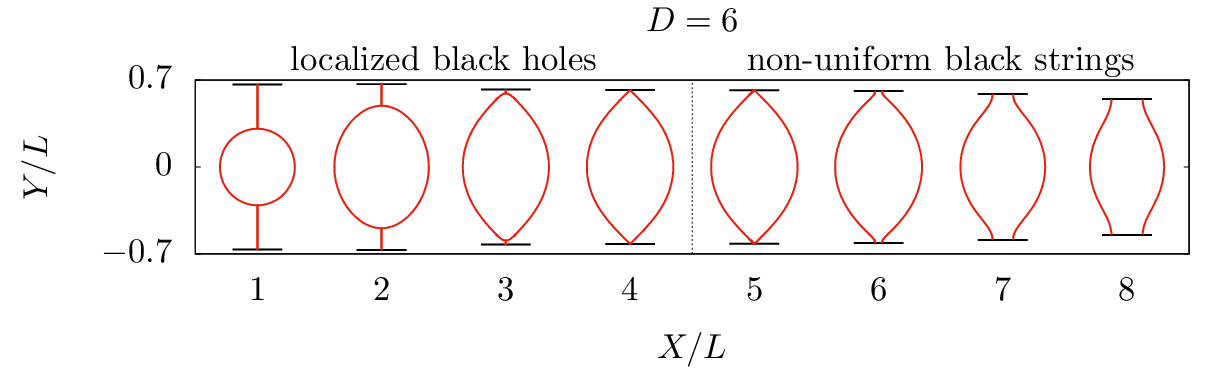}
	\caption{Spatial embedding of the horizons of different localized black hole and non-uniform black string solutions. For illustration the embeddings are shifted along the $X$-axis. The outer embeddings correspond to solutions far away from the transition, while the inner ones correspond to solutions very close to it.}
	\label{fig:Horizons}
\end{figure}

Figure~\ref{fig:Horizons} depicts the spatial embeddings of the horizon (see~\eqref{eq:embed_metric} and~\eqref{eq:embedding}) for several localized black holes and non-uniform black strings. The horizons of the localized solutions spread more and more along the compact dimension until the poles are about to touch the boundaries of the compact dimension, and hence each other due to the periodic nature of the compact dimension. From the point of view of the black string, its horizon shrinks at the boundaries of the compact dimension until it is about to pinch-off. Note that the solutions which are closest to the phase transition, corresponding to embeddings four and five in figure~\ref{fig:Horizons}, can hardly be distinguished. We refer to the region where the poles of the localized black hole are about to merge or the black string is about to pinch-off as the critical region. 

\begin{figure}
	\centering
	\includegraphics[scale=0.99]{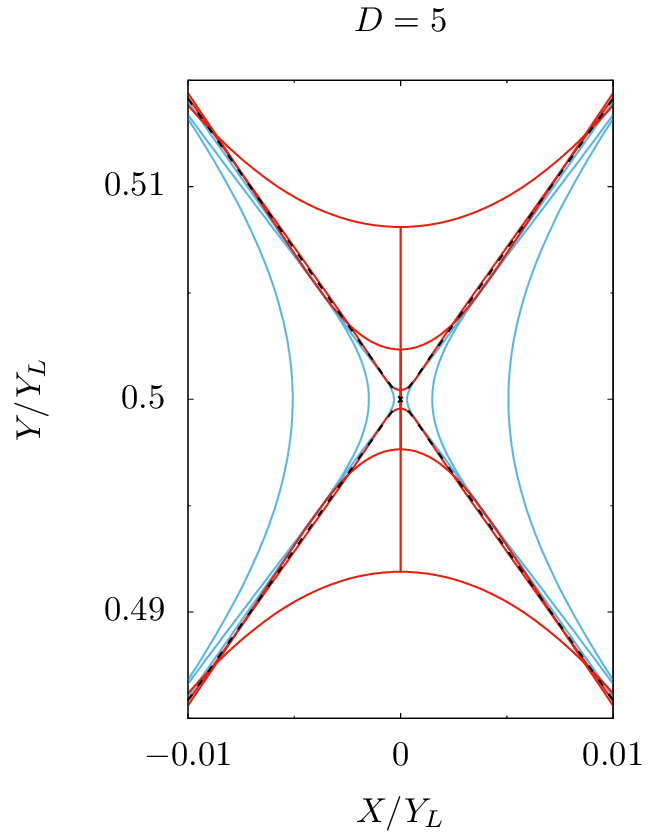} \ \\ \ \\
	\includegraphics[scale=0.99]{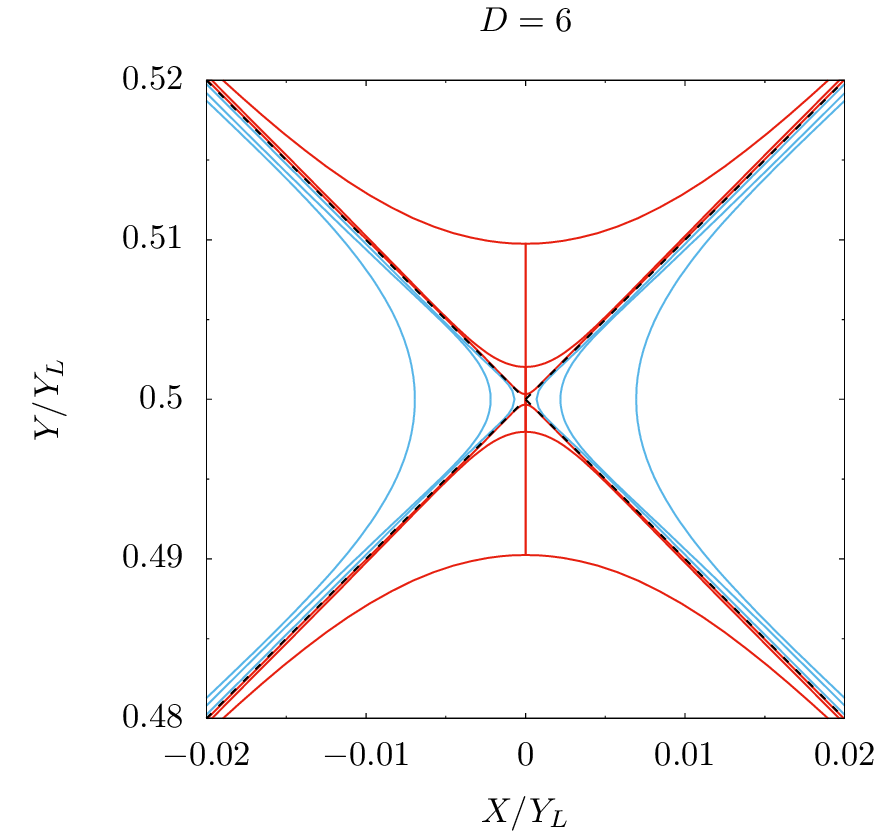}
	\caption{Magnification of the critical region where the poles of the localized black holes (red) are about to merge or the non-uniform black strings (blue) are about to pinch-off. For illustration, we have mirrored around the periodic boundary. The dashed lines correspond to the double-cone geometry, see equation~\eqref{eq:cone_embedding}. We show different localized black hole and non-uniform black string horizons that approach the cone shape from above/below and from right/left, respectively. The coordinates are normalized by $Y_L$, the length of the compact dimension as measured using the coordinate $Y$. For localized horizons, the exposed axis of symmetry is indicated connecting the poles.}
	\label{fig:Approach_Cone}
\end{figure}

In fact, we can only observe a difference between black strings and black holes in the critical region if we magnify it as in figure~\ref{fig:Approach_Cone}. Both, the localized black hole horizons and the non-uniform black string horizons, locally converge to straight lines, when approaching the transit solution.\footnote{For non-uniform black strings this was already shown in~\cite{Kalisch:2016fkm}.} This was part of the conjecture of Kol, namely that the two branches meet at a singular transit solution, which is locally given by the double-cone metric~\cite{Kol:2002xz}, see also appendix~\ref{appendix:sec:Short_review_of_the_double_cone_metric} for a short review. The predicted $D$-dependent opening angle of the horizon of the transit solution, see equation~\eqref{eq:cone_embedding}, is nicely approached from both types of solutions. However, there seems so be a qualitative difference in how the double-cone geometry is approached. In $D=6$ the double-cone geometry appears to be an envelope for both the localized black hole as the non-uniform black string horizons, whereas for $D=5$ this is only true for the black strings. The horizons of the localized solutions in $D=5$ are slightly crossing the double-cone horizon.

\subsection{Critical behavior}
\label{subsec:Critical_behavior}

Another interesting prediction, which follows from the analysis of the double-cone metric, is a critical behavior when approaching the transit solution~\cite{Kol:2002xz,Kol:2005vy}, see appendix~\ref{appendix:sec:Short_review_of_the_double_cone_metric} for a short review. In particular, different physical quantities may be expressed by the same critical exponents, which only depend on the dimension $D$. Originally, the critical behavior and their corresponding exponents appear in quantum and statistical field theories close to phase transitions. However, such a behavior was observed in some gravitational systems as well. The most famous example appears at the threshold of black hole formation in the context of a collapsing scalar field~\cite{Choptuik:1992jv}. Surprisingly, there seems to be a relation between this system in $D-1$ dimensions and the black hole/black string system in $D$ dimensions~\cite{Kol:2005vy}.\footnote{See~\cite{Sorkin:2005vz} for an investigation of the spherical collapse of a scalar field in higher dimensions.} 

In order to show such a critical behavior in the black hole/black string transition, we fit the data of different physical quantities close to the transition with an appropriate ansatz, that is
\beq
	f(Q) = f_\text{c} + a\, Q^b \cos \left(c \log Q + d \right) \, ,
	\label{eq:fit_ansatz_mass}
\eeq
with the free parameters $f_\text{c}$, $a$, $b$, $c$ and $d$, cf.\ equation~\eqref{eq:dp_perturbed_cone_real} in appendix~\ref{appendix:sec:Short_review_of_the_double_cone_metric}. The function $f$ denotes any physical quantity of interest, such as mass, entropy, temperature or relative tension. Furthermore, $f$ depends on $Q$, which uniquely parametrizes the localized black holes or non-uniform black strings. We choose $Q$ such that the critical transit solution is given by $Q=0$. For localized black holes, $Q$ is the inter-polar distance along the exposed axis, $L_\text{axis}$, and for non-uniform black strings we identify $Q$ with the minimal horizon areal radius $\Rmin$. To be more precise, we define
\refstepcounter{equation}\label{eq:Q}
\begin{align}
	Q_\text{lbh} &= L_\text{axis}/L 	 \, , \tag{\theequation a} \label{eq:Qlbh} \\
	Q_\text{nbs} &= \Rmin /R_\GL  \, , \tag{\theequation b} \label{eq:Qnbs} 
\end{align}
where $R_\GL$ is the horizon areal radius of a uniform black string at the Gregory-Laflamme point. Due to the chosen normalization, $Q$ approaches one for the starting point of the corresponding branch, i.e.\ for an infinitesimal localized black hole or a marginally stable uniform black string. 

In equation~\eqref{eq:Q} $f_\text{c}$ denotes the value of $f$ at the critical transit solution, i.e.\ for $Q\to 0$, while $b$ and $c$ are the real critical exponent and log-periodicity, respectively. Following~\cite{Kol:2002xz,Kol:2005vy} the values of $b$ and $c$ can be derived from the double-cone metric and hence should be the same for both branches and for different physical quantities $f$, see appendix~\ref{appendix:sec:Short_review_of_the_double_cone_metric}, and in particular equations~\eqref{eq:complex_exponents} and~\eqref{eq:dp_perturbed_cone_real} therein. The predicted values are $b=3/2$ and $c=\sqrt{15}/2\approx 1.9365$ for $D=5$ and $b=c=2$ for $D=6$, respectively.

\begin{figure}
	\centering
	\includegraphics[scale=0.98]{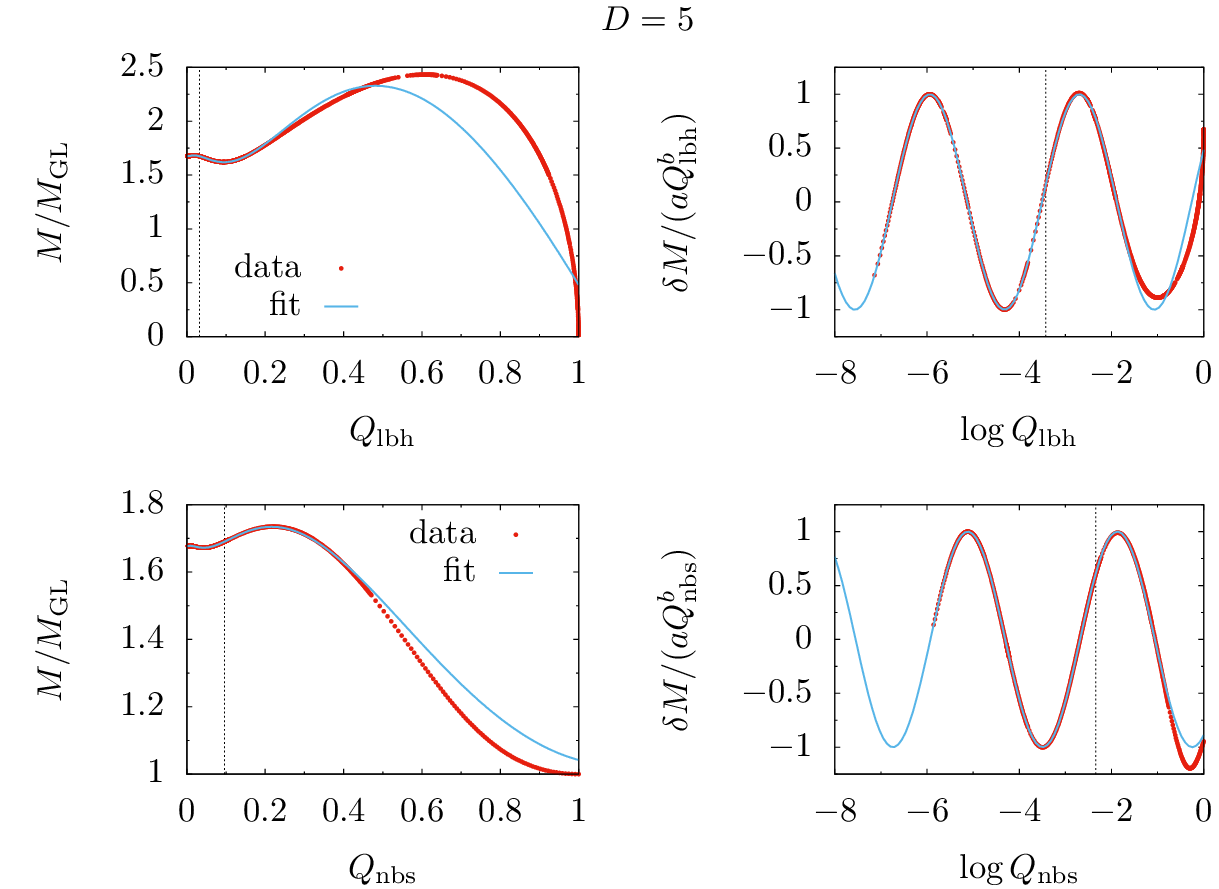} \ \\
	\includegraphics[scale=0.98]{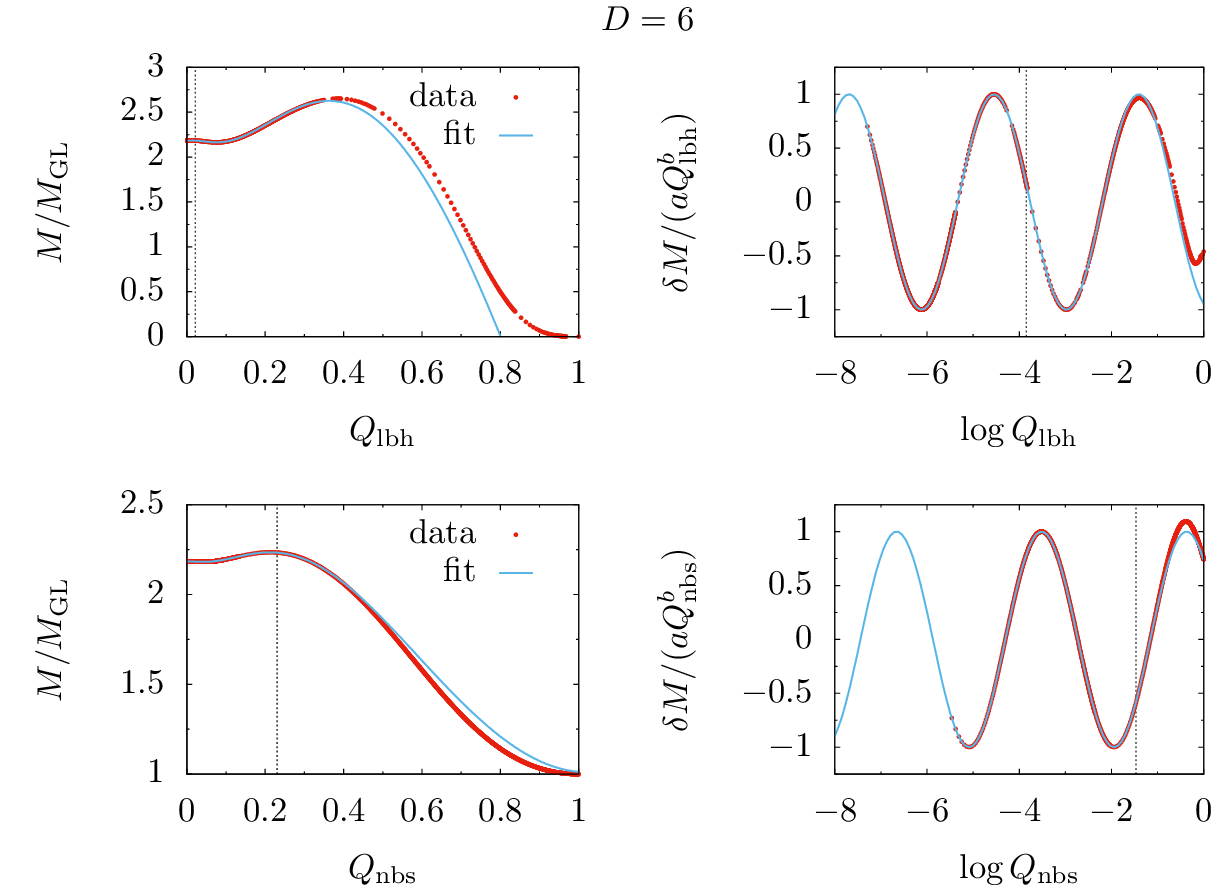}
	\caption{Data points (red dots) and fit (blue solid lines) using~\eqref{eq:fit_ansatz_mass} of the mass as a function of $Q_\text{lbh}$ or $Q_\text{nbs}$~\eqref{eq:Q}, respectively. In the left column the explicit functional dependence is shown. To resolve the tiny oscillations of the functions, a rescaled version is shown in the right column, where $\delta M = M/M_\GL -f_\text{c}$. The first two rows correspond to $D=5$ and the last two rows to $D=6$. In each plot, the dotted vertical line indicates which data points we used to produce the fit, namely all data points left to that line.  }
	\label{fig:Scaling_Mass}
\end{figure}

We used Mathematica's fit routine to obtain the values of the free parameters $f_\text{c}$, $a$, $b$, $c$ and $d$. As an example, in figure~\ref{fig:Scaling_Mass} we show data points and the corresponding fit for the mass $M$ (normalized by its value of a marginal stable uniform black string, $M_\GL$). The left column displays the actual function $M(Q)$ and we see that the fit indeed approaches the data points for small values of $Q$. Since the amplitude and the period of the oscillations in~\eqref{eq:fit_ansatz_mass} are decreasing exponentially for decreasing $Q$, a rescaling of the axes is performed in the right column of figure~\ref{fig:Scaling_Mass}. This shows more explicitly the oscillating behavior and the good agreement of data points and fit for small $Q$.

We list the numerical values for the fit parameters of mass, relative tension, entropy and temperature in table~\ref{tab:param_thermo}. The predicted values of $b$ and $c$ are excellently reproduced, with deviations of less than $0.5\%$, and, moreover, for a given dimension they are the same for different quantities and for both branches. Furthermore, the respective values $f_\text{c}$ of the critical transit solution coincide for both branches up to 7 digits after the decimal point. This is therefore the best estimation of their actual values so far.

Let us stress that not all of the data points were used for the fit. Of course, for $Q\approx 1$ the quantities do not behave like the fitting function~\eqref{eq:fit_ansatz_mass} and so only data points with reasonably small $Q$ are considered for the fitting procedure. To obtain the values listed in table~\ref{tab:param_thermo} we only take the data points of about the last cycle into account. Extending the fitting range by including more data points results in slightly bigger deviations of the parameters $b$ and $c$ from the predicted ones, as expected since the ansatz~\eqref{eq:fit_ansatz_mass} is only valid for small $Q$. The standard error of the fit routine for each parameter is of the order of the last digit (or even smaller) we gave in table~\ref{tab:param_thermo}, respectively. 
\begin{table}
	\centering
	\begin{tabular}{|c|ccccc|}
 		\multicolumn{6}{c}{$D=5$} \\
 		\hline
		\multicolumn{6}{|c|}{localized black holes} 			\\
 		\hline
				 	& $f_\text{c}$ 	& $a$	& $b$	& $c$	& $d$ 			\\
		\hline
		$M/M_\GL$ 	& 1.6771933 & 2.4700	& 1.4997 	& 1.9362 	& 2.0766 	\\
 		\hline
	 	$n/n_\GL$	& 0.7795283 & 0.5762	& 1.4986 	& 1.9359 	& 4.2842 	\\
	 	\hline
		$S/S_\GL$ 	& 2.6718298 & 7.3502	& 1.5001 	& 1.9359 	& 2.0752 	\\
		\hline
	 	$T/T_\GL$ 	& 0.6738645 & 0.7869	& 1.4990 	& 1.9367 	& 5.3444 	\\
	 	\hline
	 	\hline
		\multicolumn{6}{|c|}{non-uniform black strings} 			\\
	 	\hline
					& $f_\text{c}$ 	& $a$	& $b$	& $c$	& $d$ 			\\
		\hline
		$M/M_\GL$ 	& 1.6771932 & 0.7161 	& 1.4995 	& 1.9364 	& 3.6215	\\
		\hline
	 	$n/n_\GL$ 	& 0.7795282 & 0.1691 	& 1.5010 	& 1.9375 	& 5.8367	\\
	 	\hline
		$S/S_\GL$ 	& 2.6718297 & 2.1232 	& 1.4994 	& 1.9369 	& 3.6237	\\
		\hline
		$T/T_\GL$	& 0.6738646 & 0.2295	& 1.4998	& 1.9358 	& 0.6010 	\\
		\hline
	\end{tabular}	
	\ \\ \ \\ \ \\
	\begin{tabular}{|c|ccccc|}
 		\multicolumn{6}{c}{$D=6$} \\
 		\hline
		\multicolumn{6}{|c|}{localized black holes} 			\\
 		\hline
					& $f_\text{c}$ 	& $a$	& $b$	& $c$	& $d$ 			\\
		\hline
		$M/M_\GL$ 	& 2.1839096 & 4.75319 & 1.99999 & 1.99993 & 5.95517 \\
 		\hline
 		$n/n_\GL$ 	& 0.5855194 & 0.93638 & 1.99991 & 1.99994 & 1.70328 \\
 		\hline
		$S/S_\GL$ 	& 3.0961719 & 9.61169 & 2.00001 & 1.99992 & 5.95511 \\
 		\hline
 		$T/T_\GL$ 	& 0.7419027 & 0.65522 & 1.99991 & 1.99996 & 2.92683 \\
 		\hline
 		\hline
		\multicolumn{6}{|c|}{non-uniform black strings} 					\\
 		\hline
					& $f_\text{c}$ 	& $a$	& $b$	& $c$	& $d$ 			\\
		\hline
	 	$M/M_\GL$ 	& 2.1839096 & 1.59247 & 1.99923 & 1.99932 & 0.74457 \\
		\hline
 		$n/n_\GL$ 	& 0.5855195 & 0.30918 & 1.99487 & 1.99655 & 2.76608 \\
		\hline
 		$S/S_\GL$ 	& 3.0961720 & 3.23682 & 2.00071 & 1.99891 & 0.74332 \\
		\hline
 		$T/T_\GL$ 	& 0.7419027 & 0.21640 & 1.99512 & 2.00111 & 4.00513 \\
		\hline
	\end{tabular}
	\caption{Parameter values for mass, relative tension, entropy and temperature when fitted with~\eqref{eq:fit_ansatz_mass}. }
	\label{tab:param_thermo}
\end{table}

Using the values of table~\ref{tab:param_thermo} we are also able to perform some consistency checks. For example, we checked that the critical values of the thermodynamic quantities $f_\text{c}$ indeed satisfy Smarr's relation~\eqref{eq:Smarr} to the order of $10^{-7}$. Moreover, due to the first law of black hole thermodynamics~\eqref{eq:FirstLaw}, the extreme points of mass and entropy have to coincide. Given the ansatz~\eqref{eq:fit_ansatz_mass} this implies that the phase shift $d$ should be the same for mass and entropy which is indeed the case with deviations of less than $1\%$, see table~\ref{tab:param_thermo}. Considering only the lowest order approximation~\eqref{eq:fit_ansatz_mass} for the thermodynamic functions there are three further conditions on the parameters from Smarr's relation and the first law, and they are also satisfied to a similar accuracy.

In addition, we are able to definitely answer the question raised in~\cite{Headrick:2009pv} whether there are localized black hole solutions with positive specific heat. Reference~\cite{Headrick:2009pv} provided evidence in favor of positive specific heat close to the first maximum of mass in figure~\ref{fig:Thermo}. Note that the turning point of the mass does not coincide with the first minimum of the temperature as a function of relative tension, hence giving rise to positive $\partial M/\partial T$ between these points, in agreement with~\cite{Headrick:2009pv}.

Furthermore, we find evidence for infinitely many regions with positive specific heat. According to table~\ref{tab:param_thermo} the values of the phase shift $d$ of mass and temperature differ from each other (modulo $\pi$). Note that this is not a numerical discrepancy, e.g.\ compare with the good agreement of phase shifts of the observables mass and entropy. Thus, we conclude that in each of the tiny intervals between the corresponding turning points of mass and temperature the specific heat is positive. Of course, this argument holds also for the non-uniform black string branch.

From figure~\ref{fig:Geometry} it becomes clear that there is at least one quantity at both sides of the transition for which the ansatz~\eqref{eq:fit_ansatz_mass} is not suitable even in the critical regime. It is $L_\text{polar}$ for localized black holes and $L_\text{hor}$ for non-uniform black strings. For convenience let us denote both as the horizon length. Obviously, near the transition there are no (leading) oscillations of the horizon length. Nevertheless, we tried to fit the horizon length as a function of the parameter $Q$ by adding a non-oscillating leading term to the ansatz~\eqref{eq:fit_ansatz_mass}, that reads
\beq
	L_\text{H} (Q) = L_\text{c} - a_1 Q^{b_1} + a_2 Q^{b_2} \cos \left( c_2 \log Q + d_2 \right) \, ,
	\label{eq:fit_ansatz_L}
\eeq 
where $L_\text{H}$ represents either $L_\text{polar}$ or $L_\text{hor}$ (normalized by $L$). This ansatz has seven parameters $L_\text{c}$, $a_1$, $b_1$, $a_2$, $b_2$, $c_2$ and $d_2$. Apparently, $L_\text{c}$ is the horizon length of the critical transit solution and, of course, it should take the same value no matter from which branch the critical solution is approached. Note that $b_1 < b_2$ since by definition $Q^{b_1}$ is the leading term  in the ansatz~\eqref{eq:fit_ansatz_L} for small $Q$. 

\begin{figure}
	\centering
	\includegraphics[scale=0.98]{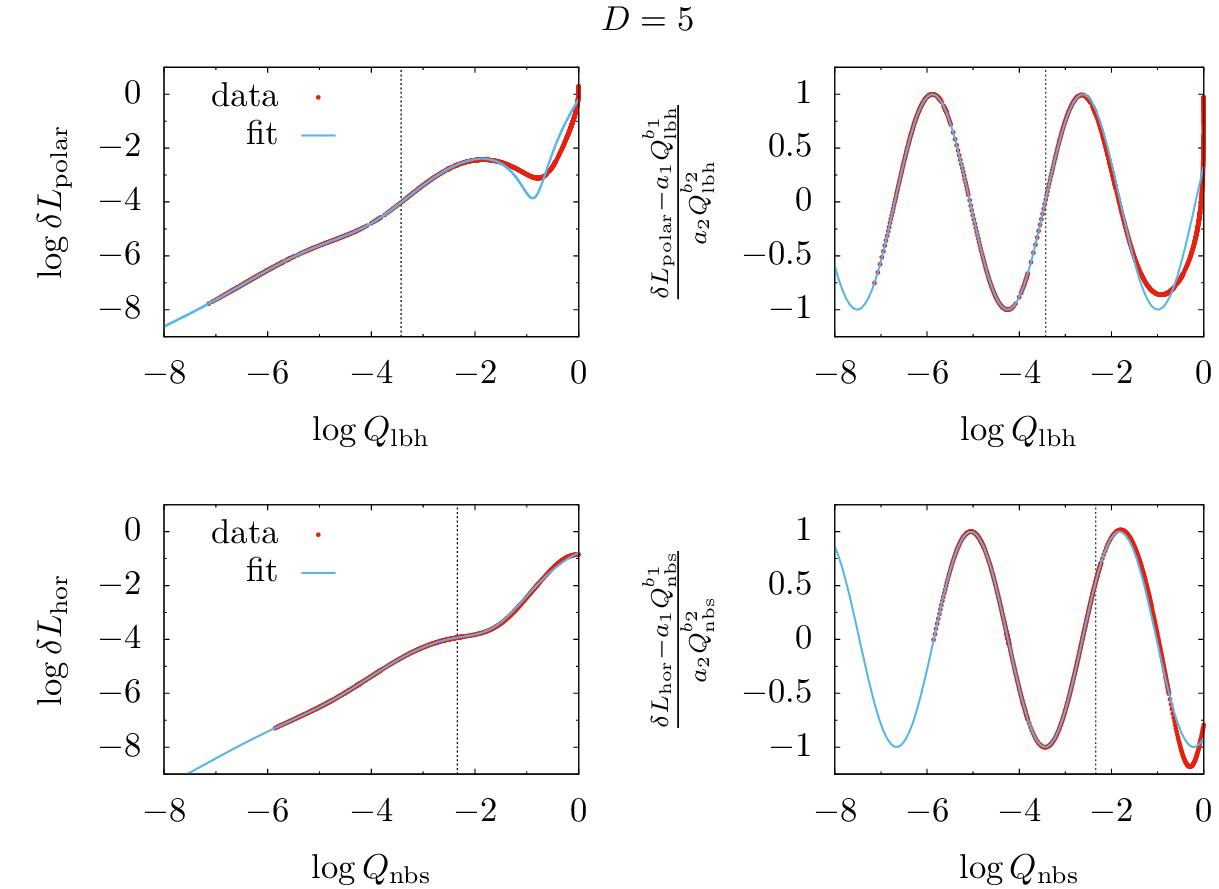} \ \\
	\includegraphics[scale=0.98]{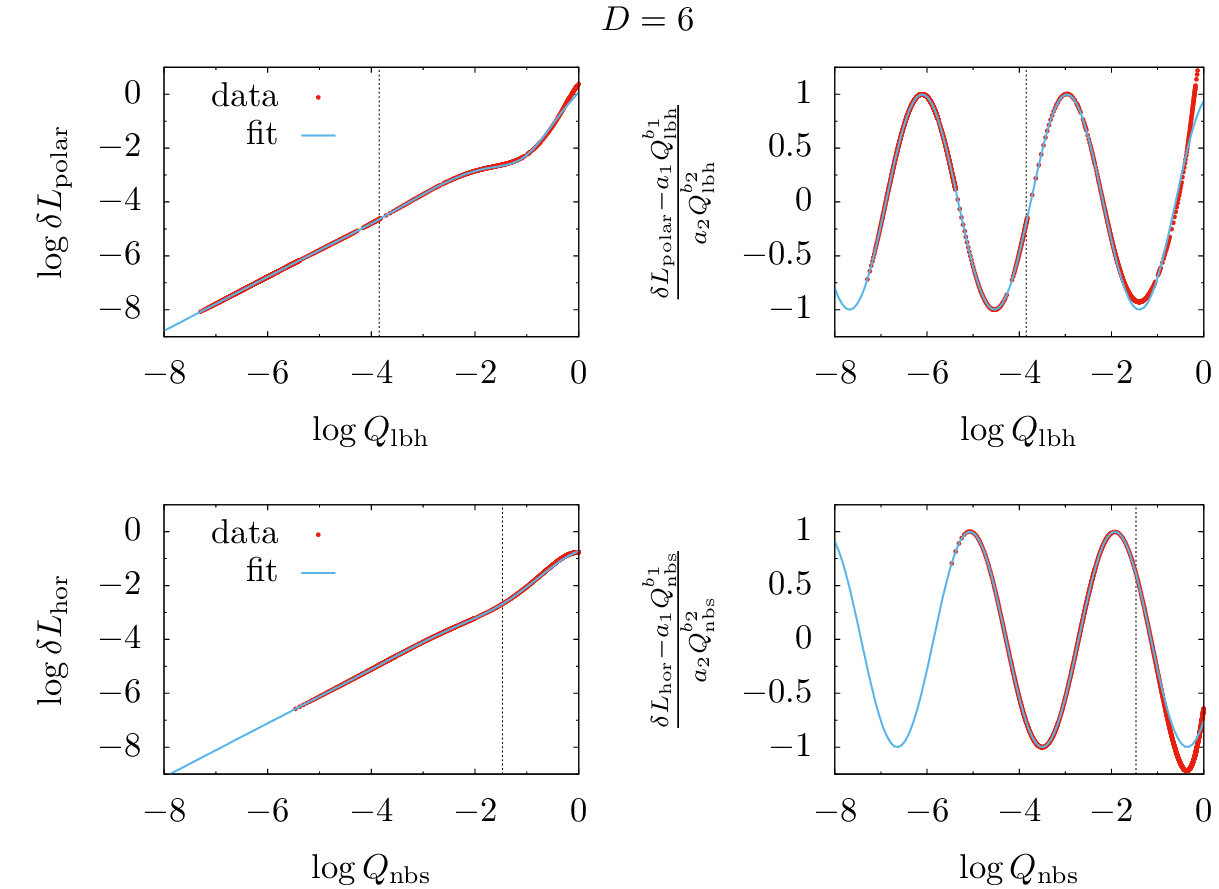}
	\caption{Data points (red dots) and fit (blue solid lines) using~\eqref{eq:fit_ansatz_L} of the horizon length $L_\text{polar}$ or $L_\text{hor}$ as a function of $Q_\text{lbh}$ or $Q_\text{nbs}$~\eqref{eq:Q}, respectively. In the left column the explicit functional dependence is shown, with both axes log-scaled. To resolve the tiny oscillations of the functions, a rescaled version is shown in the right column, where $\delta L_\text{polar} = L_\text{c} - L_\text{polar}/L$ and $\delta L_\text{hor} = L_\text{c} - L_\text{hor}/L$. The first two rows correspond to $D=5$ and the last two rows to $D=6$. In each plot, the dotted vertical line indicates which data points we used to produce the fit, namely all data points left to that line.}
	\label{fig:Scaling_HorizonLength}
\end{figure}

The resulting fit functions are again in excellent agreement with the actual behavior of the horizon length in the critical regime as evident from figure~\ref{fig:Scaling_HorizonLength}. In the left column the leading behavior is illustrated by a double logarithmic rescaling. We see that in these diagrams the functions approach straight lines, with small wiggles on top of them. These are caused by the subleading term of~\eqref{eq:fit_ansatz_L}. To show the oscillating nature of the subleading term more explicitly we rescaled the plots appropriately in the right column of figure~\ref{fig:Scaling_HorizonLength}.

We list the numerical values of the obtained fit parameters in table~\ref{tab:param_L}. Interestingly, the value of $b_1$ is approximately one for both $D=5$ and $D=6$. In other words, the horizon length is proportional to $Q$ to first order in the critical regime. Moreover, the values of $b_2$ and $c_2$ are in agreement with the critical exponents $b$ and $c$ of the thermodynamic quantities. 
\begin{table}
	\centering	
	\begin{tabular}{|c|ccccccc|}
 		\multicolumn{8}{c}{$D=5$} \\
 		\hline
					 	 	& $L_\text{c}$ 	& $a_1$		& $b_1$		& $a_2$		& $b_2$ 	& $c_2$ 	& $d_2$		\\
		\hline
		$L_\text{polar}/L$ 	& 1.428268		& 0.5548	& 1.0024	& 0.7976 	& 1.4976	& 1.9267	& 1.9116	\\
		\hline
		$L_\text{hor}/L$ 	& 1.428265		& 0.2441 	& 1.0041	& 0.2319 	& 1.5046 	& 1.9500	& 3.5614	\\
		\hline
	\end{tabular}
	\ \\ \ \\ \ \\	
	\begin{tabular}{|c|ccccccc|}
 		\multicolumn{8}{c}{$D=6$} \\
 		\hline
					 	 	& $L_\text{c}$ 	& $a_1$		& $b_1$		& $a_2$		& $b_2$ 	& $c_2$ 	& $d_2$		\\
		\hline
		$L_\text{polar}/L$ 	& 1.464800		& 0.4564 	& 0.99999	& 0.6558 	& 1.9998	& 2.0001	& 2.7909	\\
		\hline
		$L_\text{hor}/L$ 	& 1.464801		& 0.3273 	& 0.99960	& 0.2143 	& 1.9898 	& 1.9985	& 3.8530	\\
		\hline
	\end{tabular}
	\caption{Parameter values for the horizon length when fitted with~\eqref{eq:fit_ansatz_L}.}
	\label{tab:param_L}
\end{table}

\subsection{Accuracy}
\label{subsec:Accuracy}

In the previous section we have seen that the predictions for the critical exponents are confirmed with deviations of less than $0.5\%$, which provides evidence that the solutions are quite accurate even in the critical regime. However, this cannot be regarded as a measure of accuracy of the solutions itself, since only the leading behavior in the limit $Q \rightarrow 0$ is taken into account to determine the critical exponents. 

\begin{figure}
	\centering
	\includegraphics[scale=1]{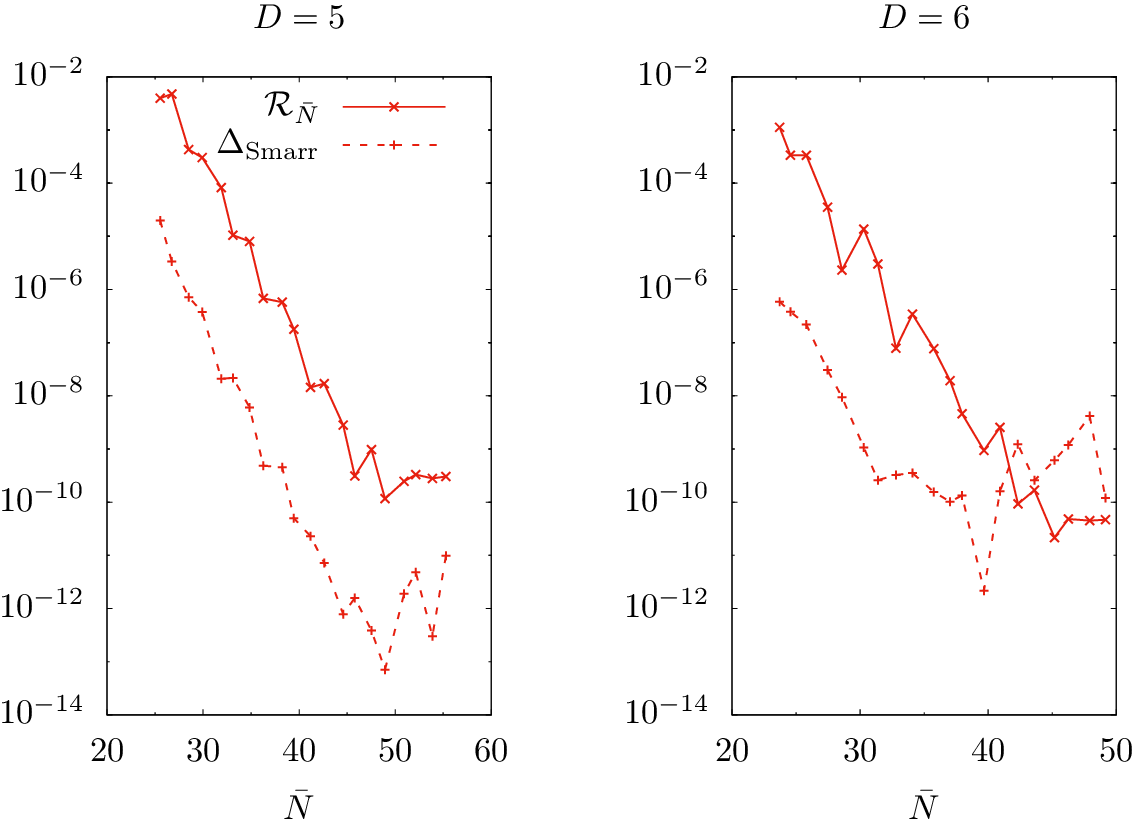}
	\caption{Convergence of the residue $\mathcal R_N$ and the deviation from Smarr's formula $\Delta _\text{Smarr}$ for localized black hole solutions close to the critical transition. The resolution $\bar N$ denotes the mean value, where we have averaged over all subdomains and all directions.}
	\label{fig:accuracy}
\end{figure}

A commonly used method to measure the accuracy of a spectral algorithm is to compare a reference solution with high resolution with solutions of lower spectral resolution, obtained by the same procedure as the reference solution. In particular, we determine all the solutions on a fine grid, using spectral interpolation techniques, and then calculate the differences of a reference solution to the solutions of lower resolution at each of these grid points. We refer to the largest magnitude of these differences as the residue $\mathcal R_N$, where $N$ indicates the resolution of the less resolved solution. By displaying $\mathcal R_N$ as a function of $N$ we observe that the residue $\mathcal R_N$ is rapidly decaying for increasing $N$, see figure~\ref{fig:accuracy} for the residue of localized black hole solutions close to the critical transition. The residue $\mathcal R_N$ saturates at values of order $10^{-9}$ (for $D=5$) or $10^{-10}$ (for $D=6$) which is due to numerical limitations caused by finite machine precision and rounding errors.\footnote{The accuracy of the non-uniform black string solutions whose data we have used in the previous sections is discussed in~\cite{Kalisch:2016fkm}.}$^,$\footnote{The numerical calculations were performed by using long double precision of the programming language C.} 

Furthermore, figure~\ref{fig:accuracy} shows the difference $\Delta _\text{Smarr}$ between right and left hand side of Smarr's formula~\eqref{eq:Smarr} for the corresponding solutions. Again, as the resolution increases, $\Delta _\text{Smarr}$ rapidly decreases until a saturation value is reached. Apparently, in $D=5$ this value is about two orders of magnitude smaller than in $D=6$. This is not surprising since in $D=6$ we have to perform two numerical derivatives at the asymptotic boundary to get the value of the coefficient $c_t$, which enters into the right hand side of Smarr's relation~\eqref{eq:Smarr}, while in $D=5$ only the first derivative is of interest, cf.\ section~\ref{subsec:Physical_quantities}.   

Finally, we note that the non-trivial components of the DeTurck vector field $\xi$ as given by~\eqref{eq:DeTurck_vector} are always smaller than $10^{-10}$ in magnitude on all grid points even for the numerical solutions closest to the critical transition. This number is negligible compared to the values of the metric functions.

\section{Conclusions}
\label{sec:Conclusions}

We constructed localized black hole solutions in five- and six-dimensional asymptotically flat Kaluza-Klein gravity with one compact periodic spatial dimension using pseudo-spectral methods. Due to the high-precision numerics and a clever choice of coordinates and domain decomposition we are able to investigate those localized black hole solutions which are about to merge into non-uniform black string solutions. 

Near the transit solution the thermodynamics associated with localized black hole solutions display a spiraling behavior that adapts to the thermodynamics of highly non-uniform black strings~\cite{Kalisch:2016fkm}.\footnote{For four-dimensional black hole systems that show a spiraling behavior see for instance references~\cite{Herdeiro:2014goa,Herdeiro:2015gia}.} We are able to resolve four turning points. Moreover, we fit the physical observables of localized black holes close to the transit solution to scaling ans\"atze with only discrete scaling symmetry. The critical exponents agree remarkably well with the ones of the conjectured double-cone metric~\cite{Kol:2002xz}, hence giving compelling evidence in favor of it. The same exponents characterize also the critical behavior of the non-uniform black string solutions. This suggests an infinite inspiral behavior of the two types of solutions. Moreover, due to this spiraling behavior and the associated phase shifts we identify infinitely many regions with positive specific heat. 

Critical behavior appears in other higher dimensional black hole configurations as well. Reference~\cite{Bhattacharyya:2010yg} showed the existence of critical exponents in the context of hairy black holes in $AdS_5\times S^5$. Several other works showed the beginning of a spiral curve in the phase diagram of hairy black holes in global $AdS_5$~\cite{Dias:2011tj} or lumpy black holes in asymptotically flat spacetime~\cite{Dias:2014cia,Emparan:2014pra}. Note that the superradiant instability of Reissner-Nordstr\" om black holes gives rise to the hairy black holes mentioned above, while the ultraspinning instability of Myers-Perry black holes leads to the lumpy black hole branch. Recall that in the black hole / black string system discussed here the non-uniform black string branch emanates from the Gregory-Laflamme instability of the uniform black strings. In all of these situations the zero-mode of the corresponding instability leads to the formation of a spiral curve in the phase diagram, which seems to be a generic feature.

Gregory-Laflamme instabilities towards non-uniform black strings and the competition between localized black holes and non-uniform black strings are generic features of Kaluza-Klein gravities irrespective of the boundary conditions such as asymptotically flat or AdS spacetimes. Hence we expect that our results and in particular our high-precision numerical methods may be useful also for other systems such as localized black holes in Anti-deSitter spacetime~\cite{Dias:2016eto,Dias:2017uyv}. In addition, the application of these methods to higher-dimensional versions of the system at hand should be straight-forward.\footnote{However, when going to higher dimensions $D>6$ at least one more technical issue will arise: the accurate extraction of the mass and the relative tension, see the discussion in appendix section~\ref{appendix:subsec:Decompositoin_of_the_domain_of_integration}.} So far, solutions for $D>6$ are rare despite the recent result for $D=10$~\cite{Dias:2017uyv}. This dimension is of particular interest for two reasons. On the one hand, using solution generating techniques, $D=10$ dimensional asymptotically Kaluza-Klein solutions correspond to type IIa supergravity solutions which are dual to certain thermal states of super-Yang-Mills theory on a circle, by the virtue of the AdS/CFT correspondence. On the other hand, according to the analysis of the double-cone metric, the critical behavior near the transit solution will change for $D\geq 10$. Hence it raises the question, whether there are oscillations of physical quantities in this regime and, consequently, whether the thermodynamics will show a spiraling behavior. 

Finally, we note that there are several other systems where the horizon topology changes and the local geometry is expected to be controlled by the double-cone metric. These are for example the transition from pinched rotating black holes to black rings in $D\geq 6$ and the transition from circular pinched rotating black holes to black saturns also in $D\geq 6$, see~\cite{Emparan:2011ve} for more information. According to the results of the work at hand we expect a critical behavior near the transition in all of these examples. In particular, the perturbative analysis of the double-cone geometry predicts the values of the critical exponents depending only on the total number of spacetime dimensions.

\section*{Acknowledgments}

We dedicate this paper to Marcus Ansorg who pushed forward the application of pseudo-spectral methods to different fields of theoretical physics including relativistic astrophysics, numerical relativity and quantum field theory. Furthermore, Marcus Ansorg utilized these methods in higher dimensional gravity which led to the work at hand.

We thank Burkhard Kleihaus and Jutta Kunz for drawing our attention to this problem. We are grateful to Barak Kol and Julian Leiber for valuable discussions. Furthermore, we thank Toby Wiseman for providing us the data of~\cite{Headrick:2009pv} for comparison. MK and SM acknowledge support by the Deutsche Forschungsgemeinschaft (DFG) graduate school GRK 1523/2.

\begin{appendix}

\section{Details on the numerical implementation}
\label{appendix:sec:Details_on_the_numerical_implementation}

The overall numerical scheme is described in section~\ref{subsec:Numerical_implementation}. In this appendix we  present more details. This involves a discussion about the reference metric functions in~\ref{appendix:subsec:Choice_of_reference_metric_functions}, and a description of the domain structure and the corresponding coordinate transformations in~\ref{appendix:subsec:Decompositoin_of_the_domain_of_integration}. Finally, in section~\ref{appendix:subsec:Parameter_values}, we discuss the parameter values that enter the numerical algorithm.

\subsection{Choice of reference metric functions}
\label{appendix:subsec:Choice_of_reference_metric_functions}

We have to specify a reference metric in order to implement the DeTurck method. As a first requirement the reference metric has to satisfy the desired asymptotic behavior and boundary conditions, see section~\ref{subsec:Metric_ansaetze_and_boundary_conditions}. For $\varrho>\varrho_{1}$ we use the flat metric~\eqref{eq:asymptotic_metric} as a reference, since it already satisfies the conditions on all boundaries except the horizon $\mathcal{H}$, see figure~\ref{fig:int_domain_bare}. In polar coordinates~\eqref{eq:polar_coordinates} the flat metric takes the form 
\beq
	\D s^2_{\mathrm{flat}} = -\D t^2 + \D \varrho ^2 + \varrho ^2 \left( \D \varphi ^2 + \sin ^2 \!\varphi \, \D \Omega ^2_{D-3} \right) \, .
	\label{eq:vac_metric_polar}
\eeq 
A good starting point for the reference metric near the horizon $\varrho = \varrho _0$ is the Schwarzschild metric, or to keep things simple
\beq
	\D s^2_{\mathrm{hor}} = -\kappa ^2(\varrho - \varrho _0)^2 \D t^2 + \D \varrho ^2 + \left[ \frac{(D-3)^2}{4\kappa ^2} + \frac{D-3}{2}(\varrho - \varrho _0)^2 \right] \D\Omega ^2_{D-2} \, ,
	\label{eq:near_horizon_metric}
\eeq
where we have used $\D \Omega ^2_{D-2} = \D \varphi ^2 + \sin ^2 \varphi \, \D \Omega ^2_{D-3}$. The metric~\eqref{eq:near_horizon_metric} approximates a Schwarzschild black hole with surface gravity $\kappa$ in the vicinity of the horizon.

We have to match these different metrics at $\varrho=\varrho_{1}$. For this purpose we write the reference metric as
\beq
	\D s^2_{\mathrm{ref}} = - H(\varrho )\, \D t^2 + \D \varrho ^2 + G(\varrho ) \, \D \Omega ^2_{D-2}
	\label{eq:ref_metric}	
\eeq
with 
\beq
	H(\varrho ) = \begin{cases}
						\lowhor H (\varrho ) & \text{if} \quad \varrho <\varrho _1 \, , \\
						1	 						 & \text{if} \quad  \varrho \geq \varrho _1 \, ,
			      \end{cases}
			      \quad \text{and} \quad
	G(\varrho ) = \begin{cases}
						\lowhor G (\varrho ) & \text{if} \quad \varrho <\varrho _1 \, , \\
						\varrho ^2  				 & \text{if} \quad  \varrho \geq \varrho _1 \, .
			      \end{cases}
	\label{eq:ref_metric_functions}
\eeq 
For $\varrho \geq \varrho _1$ we obviously recover the flat metric~\eqref{eq:vac_metric_polar}. For $\varrho <\varrho _1$ the ansatz automatically satisfies the boundary conditions on the exposed axis $\mathcal A$ and the lower mirror boundary $\mathcal M_0$, since the functions $\lowhor H$ and $\lowhor G$ do not depend on $\varphi$. The expansion of these two functions close to the horizon has to take the form
\refstepcounter{equation}\label{eq:ref_metric_functions_horizon_expansion}
\begin{align}
	\lowhor H &= \kappa ^2 (\varrho - \varrho _0 )^2 + \mathcal O \left[ (\varrho - \varrho _0 )^4 \right] \, , \tag{\theequation a} \label{eq:ref_metric_functions_horizon_expansion_a}  \\
	\lowhor G &= \frac{(D-3)^2}{4\kappa ^2} + \frac{D-3}{2}(\varrho - \varrho _0)^2 + \mathcal O \left[ (\varrho - \varrho _0 )^3 \right] \tag{\theequation b} \label{eq:ref_metric_functions_horizon_expansion_b} \, 
\end{align}
in order to satisfy the conditions on the horizon $\mathcal H$, see equation~\eqref{eq:BCh_horizon}, and to approximate the line element~\eqref{eq:near_horizon_metric}. Moreover, the functions $\lowhor H$ and $\lowhor G$ have to match the flat metric at $\varrho =\varrho _1$. 

Recall that we construct the solution numerically by means of a pseudo-spectral multi-patch scheme, i.e.\ using essentially two-dimensional splines of Chebyshev polynomials to cover the entire solution domain. Here, we choose Poincare-Steklov conditions ($\mathcal C^{1}$ continuity) at the inner domain boundaries for coupling the patches together. Based on this we now can gauge the full solution by either choosing some $\mathcal C^{\infty}$ matching of the reference metric at $\varrho=\varrho_{1}$, or some simplified $\mathcal C^{k}$ ansatz. In case of $\mathcal C^\infty$ matching we use\footnote{We incorporated this ansatz in $D=5$ and $D=6$ with the parameter values specified in section~\ref{appendix:subsec:Parameter_values}. In any case, it is crucial that the function $\lowhor G$ stays positive.}
\refstepcounter{equation}\label{eq:ref_metric_functions_horizon_exp}
\begin{align}
	\lowhor H &= 1 - E(\varrho ) \, ,  \tag{\theequation a} \label{eq:ref_metric_functions_horizon_exp_a} \\
	\lowhor G &= \varrho ^2 - E(\varrho ) \left[ \varrho ^2 - \frac{(D-3)^2}{4\kappa ^2} - (\varrho - \varrho _0)^2 \left( \frac{D^2}{4} -D +\frac{3}{4} - \kappa ^2 \varrho _0^2 \right)  \right] \, ,  \tag{\theequation b} \label{eq:ref_metric_functions_horizon_exp_b}
\end{align}
where the auxiliary function $E(\varrho)$ is given by
\beq
	E(\varrho ) = \exp \left[ -\kappa ^2 \frac{(\varrho -\varrho _0 )^2}{1-(\varrho -\varrho _0)^2 / (\varrho _1 - \varrho _0 )^2} \right] \, .
	\label{eq:exp_function}
\eeq
Obviously, the function $E(\varrho )$ is one at $\varrho =\varrho _0$ and it decays exponentially fast to zero for $\varrho \rightarrow \varrho _1$. This $\mathcal C^\infty$ matching is similar to the approach used in~\cite{Headrick:2009pv}.

In case of $\mathcal C^{k}$ matching with $k=2$, we use the following simplified ansatz
\refstepcounter{equation}\label{eq:ref_metric_functions_horizon_poly}
\begin{align}
	\lowhor H &=  \kappa ^2 (\varrho -\varrho _0)^2 + h_1 (\varrho -\varrho _0)^4 + h_2 (\varrho -\varrho _0 )^6 + h_3 (\varrho -\varrho _0 )^8 \, , \tag{\theequation a} \label{eq:ref_metric_functions_horizon_poly_a} \\
	\lowhor G &=  \frac{(D-3)^2}{4\kappa ^2} + \frac{D-3}{2}(\varrho - \varrho _0)^2  + g_1 (\varrho -\varrho _0)^4 + g_2 (\varrho -\varrho _0 )^6 + g_3 (\varrho -\varrho _0 )^8 \, . \tag{\theequation b} \label{eq:ref_metric_functions_horizon_poly_b}
\end{align}
The coefficients $h_1$, $h_2$, $h_3$, $g_1$, $g_2$ and $g_3$ can be calculated straightforwardly by matching $\lowhor H$ with $H$ and $\lowhor G$ with $G$ up to the second derivative at $\varrho =\varrho _1$. 

The $\mathcal C^{k}$ ansatz has some additional advantages. First, we do not need to take care of the essential singularity for $\varrho\rightarrow\varrho_{1}$ in contrast to the $\mathcal C^{\infty}$ ansatz. Second, the auxiliary function $E(\varrho) $ is $\mathcal C^{\infty}$ but not analytic on the regarding patch where we use the spectral approximation. This slows the convergence rate of our spectral approximation, which is circumvented by choosing polynomials for $\lowhor H$ and $\lowhor G$ which are perfectly analytic within the patches. In practice we saw no difference between the two approaches (apart from numerical fluctuations) when extracting observable quantities with the two different methods with a sufficient number of collocation points, but we saw a faster convergence of the spectral coefficients for the $\mathcal C^{k}$ ansatz.\footnote{Note that this matching procedure favors the inner boundary at the coordinate line $\varrho =\varrho _1$, see figure~\ref{fig:int_domain_all_grid}, which is now justified with hindsight.}

The $\mathcal C^{k}$ procedure rises the question about the smoothness of the target metric. First, we should keep in mind that we numerically construct a spectral spline approximation and can only monitor its convergence to the desired $\mathcal{C}^{\infty}$ solution (if existing), i.e.\ also the reference does not necessarily need to be of such a high regularity class. Furthermore, recall that the detailed form of the reference metric is a gauge choice (apart from its boundary behavior). We only need to ensure that it gives rise to some reasonable cover of the underlying manifold and is of sufficient regularity for extracting the regarding observables. 

\subsection{Decomposition of the domain of integration}
\label{appendix:subsec:Decompositoin_of_the_domain_of_integration}

\begin{figure}
	\centering
	\includegraphics[scale=1]{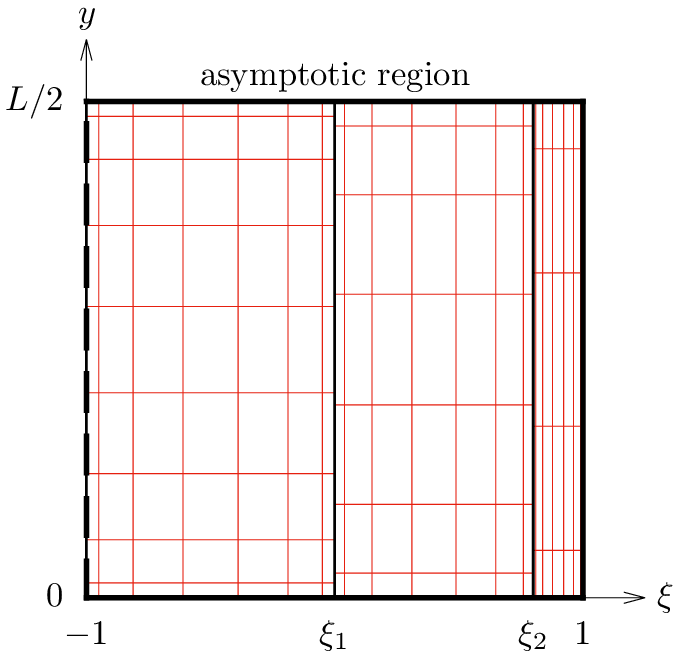}
	\caption{Domain setup in the asymptotic region with the corresponding coordinate lines. The coordinate $\xi$ compactifies infinity to the value $\xi =1$. The three subdomains are separated at $\xi = \xi _1$ and $\xi = \xi _2$. At $\xi =-1$ this region is connected to the near horizon region.}
	\label{fig:domain_decomposition_asymptotic}
\end{figure}

We introduce two different charts, the asymptotic chart~\eqref{eq:asymptotic_chart}, which is suitable to describe the spacetime in the asymptotic region $x>L/2$, and the near horizon chart~\eqref{eq:horizon_chart}, which is appropriate for the spacetime in the near horizon region $x<L/2$. The basic domain setup is shown in figure~\ref{fig:int_domain_all_grid}.

\ \\ 
\textbf{The asymptotic region} \nopagebreak \\
To cover the entire domain five (see figure~\ref{fig:int_domain_all_grid}) up to $x\to\infty$, we introduce an appropriate compactification of infinity using the new coordinate $\xi\in [-1,1]$
\beq
	x=\frac{L}{1-\xi} \, ,
	\label{eq:coordinate_xi}
\eeq
where $x=L/2$ corresponds to $\xi =-1$ and $x\to\infty$ corresponds to $\xi =1$. Near infinity we have to take care of the specific behavior of the metric functions. For $\xi\to 1$, there exist exponentially decaying $y$-dependent modes and polynomially decaying $y$-independent modes. The latter ones may even contain logarithms. The bottom line is that the fall-off of the spectral coefficients with respect to $\xi$ will be rather slow. In contrast, when $\xi$ is close to one, the fall-off of the spectral coefficients with respect to $y$ will be very rapid. Therefore, we use the following trick already employed in~\cite{Kalisch:2016fkm}: The domain five will be further divided into several (e.g.\ three) linearly connected subdomains, see figure~\ref{fig:domain_decomposition_asymptotic}. By choosing narrow domains in the vicinity of $\xi = 1$ this takes into account the non-analytic behavior of the functions. Moreover, this allows us to save a lot of grid points by adapting the resolution in $y$-direction in each of these domains. Alltogether, with this domain setup in the asymptotic region we achieve more accurate results, while simultaneously the time and memory consumption of the algorithm is reduced.

For the calculation of mass and relative tension we have to read off the asymptotic values $c_t$ and $c_y$ by differentiating the metric functions $\lowa T$ and $\lowa B$ (cf.~\eqref{eq:asymptotic_chart} and~\eqref{eq:asymptotic_corrections}) with respect to the compactified coordinate $\xi$ and then taking their values at $\xi = 1$. According to~\eqref{eq:asymptotic_corrections} we have to differentiate $(D-4)$ times. Therefore, the accuracy of $M$ and $n$ will decrease if the number of spacetime dimensions $D$ increases, since each numerical differentiation is accompanied by small errors. Nevertheless, one could naively think that it is not too difficult to avoid those errors caused by differentiation. For example the function $\lowa T$ could be expressed as $\lowa T = 1 - \tilde{\lowa T}/x^{D-4}$. The numerical scheme now could solve for the new function $\tilde{\lowa T}$ from which the value of $c_t$ can be read off directly (and without differentiation). Unfortunately, for the problem at hand things are not that simple. We refer to~\cite{Kalisch:2016fkm} where the non-uniform black string case was treated with a sophisticated ansatz for the metric functions near infinity including a detailed analysis of their asymptotic behavior, which led to highly accurate values for $M$ and $n$. As such an ansatz goes along with some subtle technical difficulties, we decided  not doing this effort in this work. However, for the results presented here, which only concern the cases $D=5$ and $D=6$, the accuracy of $M$ and $n$ is reasonably good for all constructed solutions.

\ \\ 
\textbf{The near horizon region } \nopagebreak \\
As evident from figure~\ref{fig:int_domain_all_grid} the geometries of the domains one and two favor the polar coordinates~\eqref{eq:polar_coordinates}. However, the coordinates~\eqref{eq:polar_coordinates} are not appropriate to cover the domains three and four. Therefore, we modify the polar coordinates in those two domains such that
\beq
	x = r (v,\varphi ) \sin \varphi \, , \quad  y = r ( v,\varphi ) \cos \varphi \, , 
	\label{eq:modified_polar_coordinates}
\eeq
where $v\in [\varrho _1,L/2]$ is a new radial coordinate and the function $r(v,\varphi )$ takes the form
\beq
	r(v,\varphi )= \varrho _1 \frac{L/2-v}{L/2-\varrho _1} + L/2 \frac{v-\varrho _1}{L/2-\varrho _1}
		\begin{cases}
			(\cos \varphi )^{-1} & \text{for domain 3,}  \\
			(\sin \varphi )^{-1} & \text{for domain 4.} 			
		\end{cases}
	\label{eq:modified_radial_coordinate}
\eeq
The coordinate value $v=\varrho _1$ corresponds to the contour $\varrho =\varrho _1$, while in domain 3 the coordinate value $v=L/2$ corresponds to $y=L/2$ and in domain 4 the coordinate value $v=L/2$ corresponds to $x=L/2$.

\begin{figure}
	\centering	
	\includegraphics[scale=1]{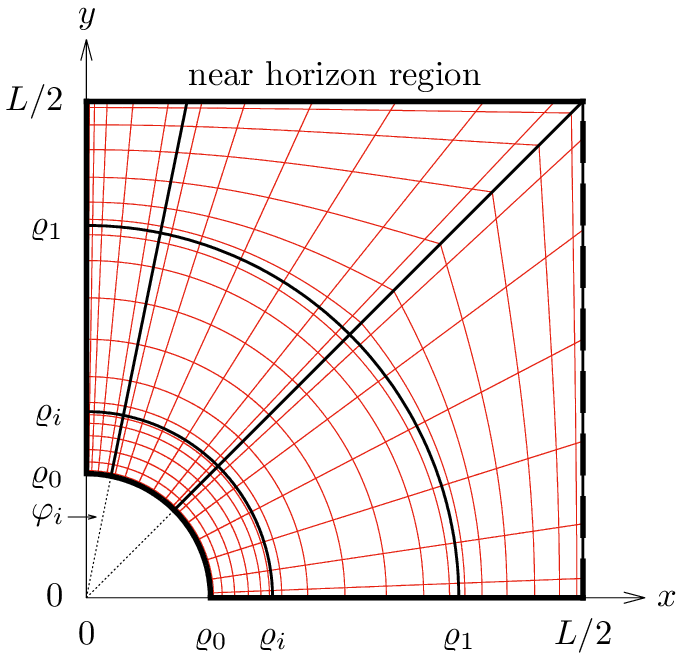}	
	\caption{Domain setup in the near horizon region with the corresponding coordinate lines. We use polar coordinates~\eqref{eq:polar_coordinates} for $\varrho \leq \varrho _1$. In the other subdomains the radial coordinate is modified according to~\eqref{eq:modified_radial_coordinate}. At $x=L/2$ this region is connected to the asymptotic region.}
	\label{fig:domain_decomposition_near_horizon}
\end{figure}

Let us now briefly discuss the functions' behavior in the near horizon region, especially near the horizon, $\varrho =\varrho _0$,  and the exposed axis, $\varphi =0$. If we consider solutions of rather tiny localized black holes, or in other words solutions with large $\kappa$ and hence low mass, the functions exhibit steep gradients near the horizon. For a better resolution in this region, we further divide the domains one and two along a contour $\varrho =\varrho _i$ with $\varrho _0<\varrho _i<\varrho _1$ into four subdomains. Consequently, we are able to increase the resolution especially in the vicinity of the horizon. Furthermore, by moving along the localized black hole branch we observed that near the exposed axis the functions develop higher and higher gradients.\footnote{This observations concerns the dimensions $D=5$ and $D=6$. We did not consider higher dimensions, but we believe that this property will be qualitatively the same.} Therefore, we use the same trick as before by choosing a $\varphi _i$ with $0<\varphi _i<\pi /4$ and by splitting all domains along this contour. Altogether, instead of the initial four subdomains in the near horizon region (see figure~\ref{fig:int_domain_all_grid}) we now have nine subdomains (see figure~\ref{fig:domain_decomposition_near_horizon}).

There are two further adaptions necessary which are important for the construction of localized black hole solutions close to the critical solution. First, while approaching this solution, all of the functions develop steep gradients near $\varphi =0$ and also the values of the functions $\lowh B$ and $\lowh S$ become exceedingly high. To ensure bounded values, we redefine the functions as 
\beq
	\lowh B = \frac{1}{\lowh{\tilde B}} \quad \text{and} \quad \lowh S = \frac{1}{\lowh{\tilde S}} \, .
	\label{eq:redefinition_Bh_and_Sh}
\eeq
Within our numerics we solve for the new functions $\lowh{\tilde B}$ and $\lowh{\tilde S}$. Second, we employ an analytic mesh refinement~\cite{meinel2012relativistic,Macedo:2014bfa} near $\varphi =0$ to flatten the occurring gradients considerably, which may lead to a much more rapid fall-off of the spectral coefficients. In our setup, we apply this trick by introducing a new azimuthal coordinate $\bar \varphi\in [0,\varphi _i]$ by 
\beq
	\varphi = \varphi _i \frac{\sinh \left( \lambda \, \bar\varphi / \varphi _i \right)}{\sinh \lambda} \, .
	\label{eq:analytic_mesh_refinement}
\eeq
As the definition suggests, this transformation is only used for $\varphi \leq \varphi _i$. The parameter $\lambda$ controls, how strong the flattening is, i.e.\ the higher $\lambda$ the stronger is the flattening. Usually, there is an optimal $\lambda$ of $\mathcal O(1)$. If this is the case, the number of coefficients to be taken to reach a certain accuracy is minimized. 

\subsection{Parameter values}
\label{appendix:subsec:Parameter_values}

We defined a set of parameters that enter the numerical scheme. Some of them have an explicit physical meaning ($L$ and $\kappa$), while the other ones are useful to fix the gauge ($\varrho _0$ and $\varrho _1$) and to control the numerical grid ($\varrho _0$, $\varrho _1$, $\varrho _i$, $\varphi _i$, $\xi _1$, $\xi _2$ and $\lambda$). In this section we provide the parameter values which were convenient for our implementation.

First, throughout the numerical calculation we chose the asymptotic circle size to be $L=8$, which sets a scale for all other parameters.\footnote{Instead, we may scale each parameter by appropriate powers of $L$ in order to obtain only dimensionless quantities. In this way, $L$ completely drops out of the numerical algorithm.} In table~\ref{tab:paravalues} we list the corresponding values of the relevant parameters. We note that these values correspond to the reference metric ansatz~\eqref{eq:ref_metric_functions_horizon_poly}, which we incorporated to approach the critical regime. 
\begin{table}
	\centering
	\begin{tabular}{|ccccccc|}
 		\hline
		 $L$	& $\varrho _0$ 	& $\varrho _1$	& $\varrho _i$	& $\varphi _i$ & $\xi _1$ & $\xi _2$		\\
		\hline
		 8 	& 0.5			& 1.5	 		& 1		 		& 0.1  			& 0		   & 0.8			\\
	 	\hline
	\end{tabular}	
	\caption{Parameter values that we chose to construct the majority of the numerical localized black hole solutions (in particular those in the critical regime) with the choice of reference metric~\eqref{eq:ref_metric_functions_horizon_poly}. We used these values both for $D=5$ and $D=6$.}
	\label{tab:paravalues}
\end{table}

However, these values are not necessarily appropriate to construct a first solution, since the corresponding initial guess (in our case  the reference metric) for the Newton method may not be close enough to the actual solution. In fact, we are rather flexible to construct a good initial guess with trial and error by fixing $L=8$ and by changing $\varrho _0$ and $\varrho _1$ accordingly. Moreover, we used   $\kappa \approx 2$ as a starting point. Note that the other parameters do not change the reference metric and hence the initial guess. Once we find a first solution, we slightly modify the surface gravity $\kappa$ to obtain another physically inequivalent solution. In this step we use the former solution as the initial guess. This procedure works fine until we reach a turning point, i.e.\ an extreme point in $\kappa$. To overcome this point we utilize the trick presented in~\cite{Dias:2015nua} section VII.B. 

Finally, we note that far from the critical regime we used small values for the parameter $\lambda$, i.e.\ $\lambda <1$, see equation~\eqref{eq:analytic_mesh_refinement}. When approaching the transition, we increased this parameter accordingly to flatten the steep gradients near the exposed axis. For our solutions which are closest to the transition we incorporated a value of $\lambda \approx 10$.

\section{Short review of the double-cone metric}
\label{appendix:sec:Short_review_of_the_double_cone_metric}

In this section we review essential features of the double-cone metric. Kol conjectured~\cite{Kol:2002xz} that the double-cone is a local model of the critical transit solution at the point where the poles of the localized black hole merge or the horizon of the non-uniform black string pinches off, respectively. The Ricci-flat metric of a cone over $\mathbb S^2\times \mathbb S^{D-3}$ reads
\beq
	\D s^2 = \D r^2 + \frac{r^2}{D-2} \left[ \D \Omega^2_2 + (D-4) \D \Omega ^2_{D-3} \right] \, .
	\label{eq:double_cone}
\eeq
Here, $r$ is the distance from the singular tip of the cone where both the $2$-sphere and the $(D-3)$-sphere have zero size. The question arises how this metric is related to the black hole/black string context. Obviously, the $\mathbb S^{D-3}$ represents the inherent spherical symmetry of the setup, while the origin of the $\mathbb S^2$ is more subtle. It is the fibration of the Euclidean time circles on a path that connects equivalent points on the horizon and crosses the periodic boundary. For example, in the localized black hole case, such a path could go along the exposed axis of spherical symmetry (see for instance figure~\ref{fig:Approach_Cone}). Clearly, as one starts at the horizon, the Euclidean time circle has zero size, while its size grows when moving away from the horizon and shrinks back to zero size when the endpoint (again on the horizon) is reached. The fibration of these circles produces a topological $\mathbb S^2$. At the merger or pinch-off point of the transit solution between localized black holes and non-uniform black strings the $\mathbb S^{D-3}$ has zero size, since the exposed axis is touched, and the $\mathbb S^2$ has zero size, since the described path has zero length. This is nicely modelled by the double-cone metric~\eqref{eq:double_cone}. See~\cite{Kol:2004ww} for a more detailed and pictorial discussion.

Geometrically, the double-cone metric dictates the shape of the horizon of the transit solution near the singular point. The embedding of this horizon into $(D-1)$-dimensional flat space~\eqref{eq:embed_metric} gives
\beq
	Y-Y_L/2 = \sqrt{\frac{2}{D-4}} \,  |X| \, ,
	\label{eq:cone_embedding}
\eeq
with an arbitrary constant $Y_L$. The tip of the cone is located at $Y=Y_L/2$. To compare with the actual localized black hole or non-uniform black string solutions, $Y_L$ is chosen to correspond to the length of the compact dimension measured in the embedding coordinate $Y$, see figure~\ref{fig:Approach_Cone}.

Furthermore, in references~\cite{Kol:2002xz,Asnin:2006ip} perturbations from the double-cone metric of the form
\beq
	\D s^2 = \D r^2 + \frac{r^2}{D-2} \left[ \E ^{\epsilon (r)} \D \Omega^2_2 + (D-4) \E ^{-2\epsilon (r) /(D-3)} \D \Omega ^2_{D-3} \right] 
	\label{eq:perturb_cone}
\eeq
were analyzed. In linear order of perturbation theory the equations of motion have the following solutions
\begin{align}
	\epsilon & = r^{s_\pm} \, , \\
	s_\pm    & = \frac{D-2}{2} \left( -1 \pm \text{i} \sqrt{\frac{8}{D-2} - 1} \right) \, .
	\label{eq:complex_exponents}
\end{align}
Obviously, for $D\geq 10$ the exponents $s_\pm$ become purely real, while for $D<10$ the imaginary part of the exponents produces oscillations in $\epsilon (r)$. 

In~\cite{Kol:2005vy} the exponents $s_\pm$ were interpreted in the following sense: Suppose $\epsilon$ is some quantity $\delta p := p-p_\text{c}$ which measures the deviation from the double-cone. Here, $p_\text{c}$ denotes the critical value of the quantity $p$ of the transit solution between localized black holes and non-uniform black strings. In addition, the coordinate $r$ is rescaled by a characteristic length scale of the perturbed cone, $r_0$. For instance, we may choose $r_0$ in such a way, that $r_0^{-2}$ is a measure of the maximal curvature of the perturbed cone. According to~\cite{Kol:2005vy} this leads to
\beq
	\delta p = \tilde A \, r_0^{-s_+} + \tilde B \, r_0^{-s_-} \, .
	\label{eq:dp_perturbed_cone}
\eeq
For $D<10$, we obtain after a short algebra
\beq
	\delta p = a \, r_0^b \cos (c\log r_0 +d) \, ,
	\label{eq:dp_perturbed_cone_real}
\eeq
with $b=-\mathfrak{Re} (s_+)$, $c=\mathfrak{Im} (s_+)$ and real constants $a$ and $d$.

\end{appendix}

\bibliographystyle{utphys}
\bibliography{LBH}

\end{document}